\documentclass[aps,pre,preprint]{revtex4-2}
\usepackage{appendix}
\usepackage{marginnote}
\usepackage{amsfonts}
\usepackage{amsmath}
\usepackage{psfrag}
\usepackage{amssymb,amsmath,amsthm}
\usepackage{color}
\newcommand{\beq}{\begin{equation}}
\newcommand{\eeq}{\end{equation}}
\newcommand{\bea}{\begin{eqnarray}}
\newcommand{\eea}{\end{eqnarray}}

\newcommand{\bma}{\left(\begin{matrix}}
\newcommand{\ema}{\end{matrix}\right)}
\newcommand{\mM}{{\mathsf M}}

\newcommand{\om}{\omega}
\begin{document}


\title{Coherent propagation and incoherent diffusion of elastic waves in a two dimensional continuum with a random distribution of edge dislocations}

\author{Dmitry Churochkin$^1$ and Fernando Lund$^2$}

\affiliation{$^1$Saratov State University, Saratov, 410012 Russia. \\
$^2$Departamento de F\'\i sica and CIMAT, Facultad de Ciencias
F\'\i sicas y Matem\'aticas, Universidad de Chile, Santiago, Chile.}


\begin{abstract}
We study the coherent propagation and incoherent diffusion of in-plane elastic waves in a two dimensional continuum populated by many, randomly placed and oriented, edge dislocations. {Because of the Peierls-Nabarro force the dislocations can oscillate around an equilibrium position with frequency $\om_0$ . The coupling between waves and dislocations is given by the Peach-Koehler force. This leads to a wave equation with an inhomogeneous term that involves a differential operator. In the coherent case, a Dyson equation for a mass operator is set up and solved to all orders in perturbation theory in independent scattering approximation (ISA). As a result, a complex index of refraction is obtained, from which an effectve wave velocity and attenuation can be read off, for both longitudinal and transverse waves. In the incoherent case a Bethe-Salpeter equation is set up, and solved to leading order in perturbation theory in the limit of low frequency and wave number. A diffusion equation is obtained and the (frequency-dependent) diffusion coefficient is explicitly calculated.  It reduces to the value obtained with energy transfer arguments at low frequency. An important intermediate step is the obtention of a Ward-Takahashi identity (WTI) for a wave equation that involves a differential operator, which is shown to be compatible with the ISA.    }
\end{abstract}

\maketitle



\section{Introduction}
{
The propagation of elastic (including acoustic) waves in complex media has evolved into a rich variety of fields and sub-fields, ranging from the very basic to the very applied. A basic tenet is that there is an elastic medium endowed with an heterogeneous structure, for example inclusions or defects. A problem that is simple to formulate is the computation of the scattering of elastic waves in an elastic medium by a static spherical inclusion that has different elastic properties. This problem is solved by use of the respective wave equation and appropriate boundary conditions at the surface of a sphere.  A next step in complexity is taken by considering not one sphere but many. Here the problem becomes unwieldy very quickly and approximation methods must be used. The resulting phenomena can be very rich, depending on the relative spacing of the spheres, the impedance contrast, and the wavelength considered. Particularly important is the geometrical location of the spheres: At one end, they can be considered to be located in a periodic lattice, and this leads to phononic crystals. At the other end, they can be randomly distributed with, say, a uniform distribution. And once randomness is allowed, there is no reason to limit it to position: size and elastic properties can also be randomly distributed. A rich variety of phenomena can ensue, depending on the relative size of wavelength, sphere radius and inter-sphere distance. For example, at appropriate wavelength resonances may occur that will considerably influence the wave propagation. And instead of spheres, other obstacle geometry can be considered. In any case, there are two major possible approaches to any one of these problems: Either the properties of the obstacles are known, and it is desired to understand the behavior of the elastic waves therein, or the behavior of the waves is known, and from their behavior it is desired to infer the properties of the heterogeneous medium.
}

{
When the obstacles to wave propagation have random intrinsic properties and/or are randomly placed, three major phenomena can occur: A coherent, average (``effective''), wave can exist. It being described by a complex index of refraction, whose real part gives an effective phase velocity, and whose imaginary part gives an attenuation. This attenuation in turn can have two origins: one, the intrinsic losses within the medium and its inclusions, and two, the loss of coherence due to the incoherent waves scattered by the random obstacles. Since the coherent wave attenuates, it will die out after a while. It may happen that the remaining, incoherent, radiation can be described in terms of a diffusion equation, with a qualitatively different wave behavior. And it may happen that the diffusion coefficient vanishes, in which case the wave can be localized (``Anderson localization'') by the disordered medium.  Experiments concerning this last phenomenon have been conducted with ultrasound traveling inside a `mesoglass', an ensemble of aluminum balls brazed together to form a disordered solid~\cite{Cobus2018}.
}

{Randomly placed elastic heterogeneities in an otherwise elastic medium can be described in two ways: On the one hand, specific obstacles can be considered, as with the spheres used for illustration in a paragraph above. In this case the randomness, so to speak, is ``discrete''. But one can also consider a medium described by a continuum random variable, for instance random, but continuously varying, elastic moduli. One object of study is the time and space evolution of the average field amplitude, also called the coherent wave. This approach was pioneered by Foldy \cite{Foldy1945} and has been widely used ever since. This coherent wave inevitably attenuates, even in the absence of inelastic losses, because of the disorder that takes, so to speak, energy out of an incoming beam. However, when this coherent wave dies out, there still remains elastic energy moving around. As emphasized by Weaver~\cite{Weaver1990}, the dynamics of this ``diffusive'' wave, is best described taking the average of the energy, that involves the average of the square of the field amplitude, rather than the square of the average of said field. In this way Weaver addressed the diffusivity of ultrasound in polycrystals. {The multiple scattering technique used by Weaver has been further elaborated upon:  For example, Thomson et al.~\cite{Thomson2008} have considered duplex microstructures; Yang et al.~\cite{Yang2012} have considered the effect of elongated grains; Li et al.~\cite{Li2015} have studied the effect of macroscopic anisotropy; and, quite recently, Bai et al.~\cite{Bai2020} have compared the behavior of two- and three-dimensional polycrystals. This methodology has been further extended to study the behavior of composites~\cite{Martin2008,Sumiya2013}. In all these works the obstacles doing the scattering are static. In the present work, as explained in detail below, we shall be concerned with a specific obstacle that responds dynamically to the incident wave.
}

{
Rather that deriving the diffusive properties of an elastic field from those of the wave amplitude, it is also possible to consider the evolution of the energy density from the beginning, bypassing the field amplitude description. In this way the specifics of wave interference arising from specific obstacles is lost, but the resuting algebra is much simpler to deal with. This approach has been describe by Ryzhik et al.~\cite{Ryzhik1996} and is widely used in seismology, where the nature of the scatterers that are obstacles to the propagation of seismic elastic waves is largely unknown~\cite{Sanchez2018}
}

{
Topological defects, such as dislocations, present an obstacle of a different nature to the propagation of elastic waves. For wavelengths longer than a few interatomic spaces, a continuum approximation is appropriate, and this is the approach we shall adopt in this work. The main difference between dislocations and other obstacles to elastic waves is that dislocations have their own dynamics: They can be treated as strings endowed with a mass and line tension, loaded by the stress associated with the elastic wave according to the well-known Peach-Koehler force~\cite{peachKoehler1950}. Their response is described by the equation of an elastic string, as pioneered by Granato and L\"ucke~\cite{Granato1956a,Granato1956b} and revisited more recently by Maurel at al. ~\cite{Maurel2005a,Maurel2005b,Natalia2009}. These strings oscillate in response to an elastic wave and, as a result, generate secondary, scattered waves. The general formalism for this aspect of the wave-dislocation interaction was discussed clearly by Mura~\cite{Mura1963} and elaborated by Lund~\cite{Lund1988}. The wave behavior in the presence of many dislocations was studied by Churochkin et al.~\cite{Churochkin2016}, who computed the coherent wave behavior, providing an expression for the (complex) index of refraction that describes such a wave. These expressions have been used to develop new non-intrusive ultrasonic non-destructive measurements of dislocation density in metals~\cite{Mujica2012,Barra2015,Salinas2017,Espinoza2018}. The quantum theory has also been studied in some detail, leading to new insights into the role played by dislocations in thermal transport~\cite{Lund2019,Lund2020}. A rather unexpected application of these results is to the understanding of the acoustic properties of glasses in the THz range~\cite{Lund2015,Bianchi2020}, where the continuum approach is helpful in getting around the fact that glasses have no crystal structure. A natural evolution of this train of thought is to explore the properties of elastic waves within a medium filled with randomly placed dislocations, in the diffusive regime. In a previous paper we have treated the case of a two-dimensional medium with screw dislocations~\cite{Churochkin2017}. In this work we tackle the problem of a two-dimensional medium as well, but populated with edge dislocations. This turns the problem from a scalar to a vector nature, and is considerably richer, as will be seen in the text that follows.
}

\subsection{Organization of this paper}
{
Section \ref{sec:2} has the basic layout of the problem to be studied: Equations in two space dimensions for the response of an edge dislocation loaded by an external, time-dependent, stress are written down, as well as the wave equation for dynamic elasticity when the source of elastic waves is a set of $N$ edge dislocations. Section \ref{sec:coh_prop} sets the problem of the propagation of waves in the presence of many dislocations as the problem of a wave in the presence of a random external potential. Standard tools are used to solve this problem to find a solution in terms of a coherent, average, wave propagating through a medium with an effective, complex and frequency-dependent, index of refraction. There appear longitudinal and transverse phase velocities, as well as longitudinal and transverse attenuations. The problem of the diffusion behavior of the waves in interaction with many edge dislocations is treated in section \ref{sec:incoh_diffu}. A Bethe-Salpeter (BS) equation is written down and a Ward-Takahashi identity (WTI) is established. As in the coherent wave problem of Section \ref{sec:coh_prop}, we use an independent scattering approximation (ISA). That is, the random variables characterizing each dislocation (in practice here, position and Burgers vector orientation) are statistically independent of each other. It is established that the WTI is consistent with the ISA. The low frequency and small wave number of said equations are considered, and the BS equation, written in the form of an eigenvalue problem, is solved perturbatively. A diffusion behavior is established, and an explicit expression for the diffusion coefficient is obtained. Section \ref{disc} has discussions and conclusions. As mentioned in the first part of this Introduction, the problem of waves interacting with topological defects that have their own dynamics has a number of different characteristics from the problem posed by static obstacles. For this reason we write down our calculations in some detail, relegating a number of the more cumbersome to several appendices. Also in an Appendix we compare the approach used here to obtain a WTI  with the approach used in \cite{Churochkin2017} for screw dislocations, and the consequences for the diffusion constant of anti-plane elastic waves traveling in a medium populated by a set of randomly located screw dislocations.
}

\section{Interaction of in-plane elastic waves with edge dislocations.}
\label{sec:2}
The problem of scattering of an in-plane wave on a single edge dislocation has been extensively treated by Maurel et al. \cite{Maurel2004}.
In turn, the coherent behavior emerges when the interaction of an in-plane wave with
an ensemble of randomly located edge dislocations is considered, as it was elucidated
by Maurel et al \cite{Maurel2004a}, using the two-dimensional adaptation of the
general equation of motion, which takes the following form:
\beq\label{wavein}
\rho \frac{\partial^2 v_a(\vec x,t)}{\partial t^2}
-c_{abcd}\frac{\partial^2}{\partial x_b\partial x_c}v_d(\vec x,t) = s_{a}(\vec x,t)
\eeq
where $\rho$ is the (two dimensional) mass density, $v_a$ is a vector of in-plane particle velocity as a function of two-dimensional position $\vec x$ and time $t$, $c_{abcd}=\lambda\delta_{ab}\delta_{cd}+\mu\left(\delta_{ac}\delta_{bd}+\delta_{ad}\delta_{bc}\right)$ is the two-dimensional elastic constants tensor with $\left(\lambda, \mu\right)$ the Lam\'{e} constants,and the source $s_{a}(\vec x,t)$ is
\beq\label{source0}
s_{a}(\vec x,t)=\sum\limits_{n=1}^{N}c_{abcd}\epsilon_{ce}\dot{X}_{e}^{n}b_{d}^{n}\frac{\partial}{\partial x_b}\delta (\vec x-\vec X^{n})
\eeq
Here the superscript $^{n}$ denotes the corresponding characteristics of the  n-th edge dislocation, {$\epsilon_{ce}=\epsilon_{ce3}$ and $\epsilon_{ijk}$ is the completely antisymmetric tensor with indexes $i,j,k, \dots=1,2,3.$},{ $\vec b$ and $\vec X$ are the Burgers vector and position of the edge dislocation, respectively, with condition $\dot{\vec{X}}\|\vec b$ (that is, an infinite straight edge dislocation line, which is directed for definiteness along $x_{3}$, moves only along its Burgers vector{, to which it is perpendicular}) and $a,b,c, \dots =1,2$}. It should be also noted that within this model the edge dislocation was originally treated as a point in two dimensions, which performs a small free oscillation without damping around a certain resting equilibrium position $\vec X_0$ (See  \cite{Maurel2004a} for details). Smallness of oscillations means that the source from Eq.[\ref{source0}] should be always taken at $\vec X=\vec X_0$, and their free character, that neglects  the Peierls-Nabarro (PN) force \cite{Nabarro1947}, leads to a scattering cross section diverging at low frequencies \cite{Maurel2004}. Qualitatively, this divergence is due to the fact that the scattering cross section must be proportional to the wavelength, which is the only length scale presented in the problem. Thus, it diverges at low frequencies when the wavelength grows. Since in the present work, a diffusion behavior associated with long wavelengths and low frequencies is of our main interest, we introduce a Peierls-Nabarro restoring force as well as a viscous damping into the edge dislocation dynamics in a similar manner as for screw dislocations in two dimensions \cite{Churochkin2017}. It yields{, for small oscillations around the PN minimum,} the following equation of motion
\beq
\label{eq_edgePN1}
M \ddot X_b  + B \dot X_b + \gamma X_b = F_b \, .
\eeq
Where $M= (\mu b^2/4\pi c_{T}^{2})\left(1+\frac{c_{T}^{4}}{c_{L}^{4}}\right) \ln (\delta/\epsilon)$ is the mass per unit length of an edge dislocation\cite{Maurel2004a} with $c_L =\sqrt{\frac{\left(\lambda+2\mu\right)}{\rho}}$ and $c_T =\sqrt{\frac{\mu}{\rho}}$; $\delta$ and $\epsilon$ are a long- and short- distance cut-off, respectively. Parameters $B$ and $\gamma$ represent a damping and the Peirls-Nabarro restoring force, correspondingly. And, finally $F_b$ is a two-dimensional version of a Peach-Koehler force $F_k=\epsilon_{kjm}\tau_m b_i\sigma_{ij}$ with $\vec \tau$ as the unit tangent vector to the dislocation line {($\hat x_3$ in our case)} and $\sigma_{ij}$ is the stress tensor taken at the dislocation line, with $\sigma_{ij}=c_{ijkl}\partial u_k/\partial x_l$, where the vector $\vec u$ is the elastic displacement{, with $\dot u_a = v_a$}.

In distinction from the original consideration of the coherent wave propagation problem  formally presented by Eqs.(\ref{wavein},\ref{source0},\ref{eq_edgePN1})(See \cite{Maurel2004a}), where an auxiliary angle $\theta$ that the {Burgers} vector makes with a specified direction was explicitly introduced, we will treat the problem in a general way, i.e. by representing the source through a combination of tensors, similarly to three-dimensional case \cite{Maurel2004}, \cite{Churochkin2016}. The natural question here is to what sort of reduction of a 3D theory our 2D case corresponds to? To answer this question, it should be recalled that in 3D, in terms of geometry, the problem is characterized by a random vector {triad} $(\vec \tau, \vec t, \vec n\equiv\vec \tau\times\vec t)$ associated with each edge dislocation, where $\vec t$ is the unit vector of the common direction of $\dot{\vec{X}}$ and $\vec b$. Reduction to 2D means fixing the equilibrium dislocation line direction, i.e. $\vec \tau$, and integrating out along the line. In other words, geometrically, it is {similar} to the Granato-Lucke configuration, {in which dislocation segments of length $L$ all point along the same direction} (See Appendix F in \cite{Maurel2005b}), resulting in two-dimensional $\vec n$ and $\vec t$

Given that $\vec \tau=(0,0,1)$, as discussed above, we can write:
\beq
\label{unitvectors}
n_a=\epsilon_{a3b}\tau_{3}t_{b}=-\epsilon_{ab}t_{b} \, .
\eeq
Then, using Eq. (\ref{unitvectors}), the following simplifications, similar to 3D case \cite{Maurel2005b}, become valid
\bea
\label{simplific}
c_{abcd}\epsilon_{ce}\dot{X}_{e}b_{d}=c_{abcd}\epsilon_{ce}\dot{X}t_{e}bt_{d}=-c_{abcd}n_{c}\dot{X}bt_{d}=-\mu \mM_{ab}\dot{X}b.
\eea
\bea
\label{simplific2}
F_at_a=\epsilon_{ab}t_ab_c c_{cbde}\partial u_d/\partial x_e=n_{b}bt_c c_{cbde}\partial u_d/\partial x_e=\mu b \mM_{ed}\partial u_d/\partial x_e.
\eea
With
\beq
\label{mmatrix}
\mM_{ab}=n_{a}t_{b}+n_{b}t_{a}\, .
\eeq
In Eq. (\ref{simplific2}) we used the explicit form for 3D Peach-Koehler force as well as the fact that the in-plane regime assumes two-dimensionality of both displacements ($\vec u$) and coordinates ($\vec x$). In addition, again smallness of vibrations means that the right-hand side of Eq. (\ref{simplific2}) should be taken at the equilibrium position of the point representing an edge dislocation, i.e. at $\vec X_0$.

Employing Eq.(\ref{simplific2}) in Eq.(\ref{eq_edgePN1}) we come to the expression for the Fourier transform of $\dot{\vec{X}}$
\bea
\label{XdotFour}
\dot X(\omega)= \frac{\mu b M_{ed}\partial v_d/\partial x_e|_{\vec x=\vec X_0}}{-\omega^2 M-Bi\omega+ \gamma}
\eea
Eventually, from Eqs. (\ref{source0},\ref{simplific},\ref{XdotFour}) one can get for the Fourier transform of 2D source
\beq\label{source0Four}
s_{a}(\vec x,\omega)=\sum\limits_{n=1}^{N}\mu{\cal A}^{n}\mM^n_{ab}\frac{\partial}{\partial x_b}\delta (\vec x-\vec X_{0}^{n})\mM^n_{ed}\frac{\partial}{\partial x_e}|_{\vec x=\vec X_0^{n}}v_d
\eeq
and
\beq\label{prefin}
{\cal A}^{n}=\frac{\mu b^2}{M} \frac{1}{\omega^2 {+} i \omega (B/M) -\omega^2_0 }
\eeq
where $\omega_0^2 \equiv \gamma /M$. {Notice that this coefficient is the same for all dislocations, since it depends on the magnitude, not the orientation, of the Burgers vector. Accordingly, we shall drop its superscript ``$n$''. The tensor $\mM_{ab}^n$, however, does depend on Burgers vector orientation and is randomly distributed over the various dislocations.} Then, the 2D version of the tensorial perturbation potential for a single edge dislocation  can be introduced in the way similar to 3D case \cite{Maurel2005b},\cite{Churochkin2016}
\beq\label{potentialFour}
V^{n}_{ad}(\vec x,\omega)=\mu{\cal A}\mM^{n}_{ab}\frac{\partial}{\partial x_b}\delta (\vec x-\vec X_{0}^{n})\mM^{n}_{ed}\frac{\partial}{\partial x_e}|_{\vec x=\vec X_0^{n}}
\eeq
And the total perturbation potential is written as a superposition
\beq\label{TotpotentialFour}
V_{ad}(\vec x,\omega)=\sum\limits_{n=1}^{N}V^{n}_{ad}(\vec x,\omega)
\eeq
It is worth mentioning that the 2D tensor from Eq. (\ref{potentialFour}) can be obtained from the 3D potential \cite{Churochkin2016}  by making use of the following formal reduction:
\bea\label{reduction}
i,j,k, \dots.&\Rightarrow& a,b,c, \dots\\
{\cal A}&\Rightarrow&\mu{\cal A}\nonumber
\eea
In what follows, we consider that all edge dislocations have a Burgers vector of the same magnitude, but randomly oriented in 2D.

\section{Coherent propagation}
\label{sec:coh_prop}
The coherent behavior of the elastic wave in the presence of many edge dislocations is described by the average Green function $\langle G \rangle$, where $G$, according to \cite{Maurel2004a} and Eqs. (\ref{wavein}, \ref{source0Four},\ref{potentialFour},\ref{TotpotentialFour}), should be the solution of
\beq
\rho\omega^2 G_{ae}(\vec x,\omega)+c_{abcd}\frac{\partial^2}{\partial x_b\partial x_c}G_{de}(\vec x,\omega) = -V_{ad}(\vec x,\omega)G_{de}(\vec x,\omega)-\delta_{ae}\delta(\vec x).
\label{greeneq}
\eeq
and the brackets denote an average over the random distribution {(of both position and Burgers vector orientation)} of edge dislocations in 2D.

The average Green function is expressed as the solution of the Dyson equation in 2D
\beq
 \langle G\rangle = \left[ (G^0)^{-1}-\Sigma \right]^{-1}
\label{Dyson}
\eeq
where $G^0$ is the Green's function of the medium without dislocations (i.e., ``free'' {or ``bare'' }) \cite{Maurel2004a,Maurel2005b,Weaver1990}. For an infinite, homogeneous and isotropic 2D medium it obeys the equation in Fourier space
\beq
\rho\omega^2 G^0_{ae}(\vec x,\omega)+c_{abcd}\frac{\partial^2}{\partial x_b\partial x_c}G^0_{de}(\vec x,\omega) = -\delta_{ae}\delta(\vec x).
\label{g0eq}
\eeq
giving as a solution
\beq
[(G^0)^{-1}]_{ab}(\vec k,\omega)= (-\rho \omega^2+\mu k^2)\delta_{ab}+(\lambda+\mu)k_ak_b
\label{green0}
\eeq
or
\begin{equation}
{G^0_{ab}}(\vec k,\omega)= \frac{1}{\rho c_T^2(k^2-  k_T^2)}(\delta_{ab}-\hat k_a \hat k_b)+\frac{1}{\rho c_L^2(k^2-k_L^2)} \hat k_a \hat k_b \, .
  \label{g0Sol}
\end{equation}
where $\vec k$ is the unperturbed wave number of the 2D medium described by $G^0$, $k_T \equiv \omega/c_T$, $k_L \equiv \omega/c_L$, and with an implied small imaginary part in the denominator to insure causality. We note that, as can be easily verified, the result in Eq. (\ref{g0Sol}) agrees up to a prefactor with that presented in \cite{Maurel2004a}. The corresponding expression in coordinate space is defined by
\beq
G^0_{ab}(\vec x -\vec x', \omega) =\frac{1}{(2\pi)^2}\int d\vec k e^{-i\vec k \cdot \vec x} {G^0_{ab}}(\vec k, \omega) e^{i\vec k \cdot \vec x'}.
\label{g0}
\eeq
Finally,  the mass (or self-energy) operator $\Sigma$, is given by
\beq
\label{MO}
\Sigma = \langle T \rangle - \langle T \rangle G^0 \Sigma
\eeq
with $T$ as the T-matrix, represented in terms of {$V$} {(given by Eqn. (\ref{TotpotentialFour})} by
\beq
\label{Tmatrix}
T = {V} + {V}G^0 T \, .
\eeq
In turn, the last equation can be easily transformed into a series in respect of a perturbation potential
\beq
T = {V} + {V}G^0{V} + {V}G^0{V}G^0{V} + \cdots
\eeq

A statistical independence of the scatterers of each other is assumed in this paper, that is an Independent Scattering Approximation (ISA) is used to treat the coherent wave propagation problem analytically as it was done in 3D \cite{Churochkin2016}. {Assuming a uniform dislocation distribution through a surface $S$} it can be easily shown that the self-energy $\Sigma$ within the ISA is evaluated as
\beq
\Sigma = \frac NS \int \langle t^n \rangle d\vec X_0^n
\label{massISA}
\eeq
where $t^n$ is the T-matrix for scattering by a single object, similar to that defined in \cite{Churochkin2016}, for 3D,  {and, in this instance, the brackets denote an average over the Burgers vector orientation}. Maurel et al. \cite{Maurel2004a} computed the mass operator to second order in perturbation theory. It is shown in \ref{Apb} that, due to the point-like nature of the interaction (\ref{TotpotentialFour}) the perturbation series is geometric and can be summed to all orders to obtain the following expression for the Green function ${<G^{+}>}({\bf k},\omega)$ for the outgoing waves in the momentum space
{
\beq
\label{green}
{<G^{+}>_{ab}}({\bf k},\omega)=\frac{(\delta_{ab}-\hat k_a \hat k_b)}{\rho \omega^{2}\left\{\frac{k^2}{K_{T}^{2}}-1\right\}}+\frac{\hat k_a \hat k_b}{\rho \omega^{2}\left\{\frac{k^2}{K_{L}^{2}}-1\right\}}
\eeq
 {with}
\begin{eqnarray}\label{poles}
K_{T, L} &=& \frac{\omega}{c_{T, L}\sqrt{\left[1-\frac{\Sigma}{\rho c_{T, L}^{2}k^{2}}\right]}}= \frac{\omega}{c_{T, L}\sqrt{1+n\left(\frac{c_{T}}{c_{T, L}}\right)^{2}{\cal T}} }\\
\Sigma_{ab} & = & \Sigma\delta_{ab} = -n\mu {\cal T} k^2\delta_{ab} \label{massop} \\
{\cal T} & = & \frac{1}{2}\frac{{\mathcal A}}{1+\mu{\mathcal A}I}
\label{eq:tee}
\end{eqnarray}
Here, we note that the mass operator from Eq.(\ref{massop}) suggests a mode insensitive effect (See Eq.(\ref{poles})) of an ensemble of randomly distributed vibrating edge dislocations on the effective wave numbers $K_{T, L}$ of the coherently propagating in-plane elastic waves irrespectively  of their types. In this regards, the problem under consideration is expected to be similar to that, which has been recently formulated and treated in \cite{Churochkin2017} for a purely scalar case of screw dislocations in two dimensions. On the other side, the two-dimensional character of the problem is still held and exhibited through a tensorial structure of the mass operator,
which makes the Green's tensor from Eq. (\ref{green}) resembling the one characteristic for the three-dimensional problem \cite{Churochkin2016}.
The above mentioned similarities are  extensively used further when the problem of diffusion  of the elastic waves is treated.}

\section{Incoherent diffusion}
\label{sec:incoh_diffu}
{We highlighted above an intermediate character (bearing scalar and tensorial attributes) of the problem under consideration to proceed with the incoherent (diffusive) transport of the elastic waves. Indeed, we naturally adopt some aspects of a scalar version of the spectral approach, that was applied to a medium with random scatterers of a screw dislocation type (See, \cite{Churochkin2017} and references therein for details), and generalize it to our 2d tensorial case with randomly distributed vibrating edge dislocations. Meanwhile, to save the space we use bold notations for 2d tensors or vectors from now on unless indices are explicitly required to avoid misunderstanding.}

\subsection{Bethe-Salpeter {(BS)} equation and Ward-Takahashi Identity}
\label{BSWT}
\subsubsection{BS equation}
{Following \cite{Churochkin2017}, to describe the diffusive transport regime, the two-point correlation  $<\bf{G}^{+}\otimes\bf{G}^{-}>$ needs to be determined in the low frequency,
long wavelength limit; the operation $\otimes$ is defined for arbitrary second rank tensors $\bf{A}$ and $\bf{B}$ as: ($\bf{A}\otimes \bf{B}$)$_{ab,cd}$=$A_{ac}B_{db}$. To this end, in Fourier space, a diffusive pole structure, $<\bf{G}^{+}\bf{G}^{-}>\, \sim (\imath \Omega +q_{a}D_{ab}q_{b})^{-1}$ should be found, with $\bf{D}$ identified as a diffusion tensor ~\cite{Sheng2006}. The correlation,
for its part, obeys the Bethe-Salpeter (BS) equation, which can be constructed in two steps:
First, $<\bf{G}^{+}\otimes\bf{G}^{-}>$ is formally rewritten as
\begin{small}
\begin{eqnarray}\label{Intensity}
<\mathbf{G^{+}}\otimes\mathbf{G^{-}}>&=&<\mathbf{G^{+}}>\otimes<\mathbf{G^{-}}>+
\left(<\mathbf{G^{+}}\otimes\mathbf{G^{-}}>-<\mathbf{G^{+}}>\otimes<\mathbf{G^{-}}>\right)\\
&=&<\mathbf{G^{+}}>\otimes<\mathbf{G^{-}}>  +<\mathbf{G^{+}}>\otimes<\mathbf{G^{-}}>:<\mathbf{G^{+}}>^{-1}\otimes<\mathbf{G^{-}}>^{-1}:\nonumber\\
&&\left(<\mathbf{G^{+}}\otimes\mathbf{G^{-}}>-<\mathbf{G^{+}}>\otimes<\mathbf{G^{-}}>\right):
<\mathbf{G^{+}}\otimes\mathbf{G^{-}}>^{-1}:\nonumber\\
&&<\mathbf{G^{+}}\otimes\mathbf{G^{-}}>\nonumber\\
&=&<\mathbf{G^{+}}>\otimes<\mathbf{G^{-}}>+<\mathbf{G^{+}}>\otimes<\mathbf{G^{-}}>:\nonumber \\
&&\left(<\mathbf{G^{+}}>^{-1}\otimes<\mathbf{G^{-}}>^{-1}-<\mathbf{G^{+}}\otimes\mathbf{G^{-}}>^{-1}\right)
:<\mathbf{G^{+}}\otimes\mathbf{G^{-}}>\nonumber
\end{eqnarray}
\end{small}
where {the tensor scalar product} $:$ is defined as {$[(\bf{A}\otimes \bf{B}$):($\bf{C}\otimes \bf{D}$)]$_{ab,ef}$=($\bf{A}\otimes \bf{B}$)$_{ab,cd}$($\bf{C}\otimes \bf{D}$)$_{cd,ef}$. Note that $[(A \otimes B):(A^{-1} \otimes B^{-1})]_{ab,ef} = \delta_{ae} \delta_{{fb}}$}.
Second, the pole structure can be easily introduced from the last expression of the Eqn. (\ref{Intensity}) by defining the irreducible {vertex} $\mathbf{K}$ as
\begin{eqnarray}\label{Irred}
\mathbf{K} \, {\equiv}<\mathbf{G^{+}}>^{-1}\otimes<\mathbf{G^{-}}>^{-1}-<\mathbf{G^{+}}\otimes\mathbf{G^{-}}>^{-1}
\end{eqnarray}
and substituting it into Eqn. (\ref{Intensity}), which result in the BS equation of the well known form~\cite{Sheng2006}
\beq\label{IntensityForm}
<\mathbf{G^{+}}\otimes\mathbf{G^{-}}>=
<\mathbf{G^{+}}>\otimes<\mathbf{G^{-}}>+\\
<\mathbf{G^{+}}>\otimes<\mathbf{G^{-}}>:\mathbf{K}:<\mathbf{G^{+}}\otimes\mathbf{G^{-}}> \, ,
\eeq
where $\mathbf{K}$ plays a role similar to the role played by the self-energy $\Sigma$ in Dyson's equation, Eqn. (\ref{Dyson}).
If we define the {two-sided} Fourier transforms as\cite{Sheng2006}
\begin{eqnarray}\label{BSft}
\mathbf{G}^{+}(\mathbf{x}_{1},\mathbf{x}^{\prime}_{1};\omega^{+})=\int\limits_{\mathbf{k}_{1}} \int\limits_{\mathbf{k}^{\prime}_{1}}e^{\imath \mathbf{k}_{1}\mathbf{x}_{1}} \mathbf{G}^{+}(\mathbf{k}_{1},\mathbf{k}^{\prime}_{1};\omega^{+})e^{-\imath \mathbf{k}^{\prime}_{1}\mathbf{x}^{\prime}_{1}}\\
\mathbf{G}^{-}(\mathbf{x}_{2},\mathbf{x}^{\prime}_{2};\omega^{-})=\int\limits_{\mathbf{k}_{2}} \int\limits_{\mathbf{k}^{\prime}_{2}}e^{-\imath \mathbf{k}_{2}\mathbf{x}_{2}} \mathbf{G}^{-}(\mathbf{k}^{\prime}_{2},\mathbf{k}_{2};\omega^{-})e^{\imath \mathbf{k}^{\prime}_{2}\mathbf{x}^{\prime}_{2}} \nonumber
\end{eqnarray}
then
\beq
\label{ApBSft}
<\mathbf{G^{+}}\otimes\mathbf{G^{-}}>={\int d\left(\frac{\mathbf{R}+\mathbf{R}^{\prime}}{2}\right)\mathbf{G^{+}}\otimes\mathbf{G^{-}}=}
\int\limits_{\mathbf{k}} \int\limits_{\mathbf{k}^{\prime}}\int\limits_{\mathbf{q}}\mathbf{\Phi}({\bf k},{\bf k}^{\prime};{\bf q},\Omega)
e^{\imath\left(\mathbf{k}\mathbf{r}-\mathbf{k}^{\prime}\mathbf{r}^{\prime}+\mathbf{q}\left(\mathbf{R}-\mathbf{R}^{\prime}\right)\right)}
\eeq
where
\begin{eqnarray}\label{intensity}
\mathbf{\Phi}({\bf k},{\bf k}^{\prime};{\bf q},\Omega)\equiv <\mathbf{G}^{+}(\mathbf{k}^{+},\mathbf{k}^{\prime +},\omega^{+})\otimes\mathbf{G}^{-}(\mathbf{k}^{\prime-},\mathbf{k}^{-},\omega^{-})>
\end{eqnarray}
 {with}
\beq
\mathbf{k}^{\pm}=\mathbf{k}\pm\frac{\mathbf{q}}{2},\quad\omega^{\pm}=\omega\pm\frac{\Omega}{2} \, .\nonumber
\eeq
And space variables have been introduced as
\begin{eqnarray}\label{BSspace}
\mathbf{x}_{1}=\mathbf{R}+\frac{\mathbf{r}}{2}, & &
\mathbf{x}_{2}=\mathbf{R}-\frac{\mathbf{r}}{2}\\
\mathbf{x}^{\prime}_{1}=\mathbf{R}^{\prime}+\frac{\mathbf{r}^{\prime}}{2}, & &
\mathbf{x}^{\prime}_{2}=\mathbf{R}^{\prime}-\frac{\mathbf{r}^{\prime}}{2}\nonumber\\
\mathbf{k}_{1}=\mathbf{k}^{+}=\mathbf{k}+\frac{\mathbf{q}}{2}, & &
\mathbf{k}^{\prime}_{1}=\mathbf{k}^{\prime+}=\mathbf{k}^{\prime}+\frac{\mathbf{q}}{2}\nonumber\\
\mathbf{k}^{\prime}_{2}=\mathbf{k}^{\prime-}=\mathbf{k}^{\prime}-\frac{\mathbf{q}}{2}, & & \mathbf{k}_{2}=\mathbf{k}^{-}=\mathbf{k}-\frac{\mathbf{q}}{2} \, .\nonumber
\end{eqnarray}

By applying the inverse Fourier transform\cite{Stark1997}
\begin{eqnarray}\label{ApBSIft}
\int d\left(\mathbf{R}-\mathbf{R}^{\prime}\right)d\mathbf{r}d\mathbf{r}^{\prime}
e^{-\imath\left(\mathbf{k}\mathbf{r}-\mathbf{k}^{\prime}\mathbf{r}^{\prime}+\mathbf{q}\left(\mathbf{R}-\mathbf{R}^{\prime}\right)\right)} \end{eqnarray}
to Eqn. (\ref{IntensityForm}) the BS equation in momentum space is obtained as
\bea
\label{BSmom}
\mathbf{\Phi}({\bf k},{\bf k}^{\prime};{\bf q},\Omega)&=&<\mathbf{G}^{+}>\otimes<\mathbf{G}^{-}>({\bf k};{\bf q},\Omega)\delta_{{\bf k},{\bf k}^{\prime}}\\
&& \hspace{-2em} +\int\limits_{\mathbf{k}^{\prime\prime}}<\mathbf{G}^{+}>\otimes<\mathbf{G}^{-}>({\bf k};{\bf q},\Omega):\mathbf{K}({\bf k},{\bf k}^{\prime\prime};{\bf q},\Omega):\mathbf{\Phi}({\bf k}^{\prime\prime},{\bf k}^{\prime};{\bf q},\Omega)\nonumber
\eea
with $\delta_{{\bf k},{\bf k}^{\prime}}=(2\pi)^{2}\delta({\bf k}-{\bf k}^{\prime})$ and, as usual, the integration over the internal momentum variables, i.e. ${\bf k}^{\prime\prime}$, is assumed, with $\int\limits_{\mathbf{k}} \equiv \frac{1}{(2\pi)^{2}}\int d\mathbf{k}$.
Proceeding from Eqn. (\ref{BSmom}) to a kinetic form requires using the following auxiliary identity for the  averaged Green's functions:
\beq
\label{BSidentity}
(<\mathbf{G^{+}}>^{-1}\otimes\mathbf{I}-\mathbf{I}\otimes<\mathbf{G^{-}}>^{-1}):(<\mathbf{G}^{+}>\otimes<\mathbf{G}^{-}>)=   (\mathbf{I}\otimes<\mathbf{G^{-}}>-<\mathbf{G^{+}}>\otimes\mathbf{I})
\eeq
with $\mathbf{I}$ as a unit tensor.
{ Now, acting with $(<\mathbf{G^{+}}>^{-1}\otimes\mathbf{I}-\mathbf{I}\otimes<\mathbf{G^{-}}>^{-1})$ on the left of Eqn. (\ref{BSmom}) and using Eqn. (\ref{BSidentity}) it is straightforward to obtain
\bea
\label{BSkinetic}
(<\mathbf{G^{+}}>^{-1}\otimes\mathbf{I}-\mathbf{I}\otimes<\mathbf{G^{-}}>^{-1}):\mathbf{\Phi}&=&\\
&& \hspace{-8em}(\mathbf{I}\otimes<\mathbf{G^{-}}>-<\mathbf{G^{+}}>\otimes\mathbf{I}):
\left(
\mathbf{I}\otimes\mathbf{I}\delta_{{\bf k},{\bf k}^{\prime}}+
\mathbf{K}({\bf k},{\bf k}^{\prime\prime};{\bf q},\Omega):\mathbf{\Phi}({\bf k}^{\prime\prime},{\bf k}^{\prime};{\bf q},\Omega)\right)\nonumber
\eea
}
and, using Dyson's equation Eqn. (\ref{Dyson}) and the free space Green tensor Eqn. (\ref{g0Sol}) this can be eventually reduced to
the kinetic form of the BS equation
\beq
\label{BSfinal}
\left[\imath\omega\Omega\mathbf{E}+\mathbf{P}({\bf k};{\bf q})\right]:\mathbf{\Phi}({\bf k},{\bf k}^{\prime};{\bf q},\Omega)
+\int\limits_{\bf{k}^{\prime\prime}}\mathbf{U}({\bf k},{\bf k}^{\prime\prime};{\bf q},\Omega):\mathbf{\Phi}({\bf k}^{\prime\prime},{\bf k}^{\prime};{\bf q},\Omega)
=\delta_{{\bf k},{\bf k}^{\prime}}\mathbf{\Delta G}({\bf k};{\bf q},\Omega)
\eeq
with
\begin{eqnarray}\label{potBS}
\mathbf{U}({\bf k},{\bf k}^{\prime};{\bf q},\Omega) \equiv
\Delta\mathbf{\Sigma}({\bf k};{\bf q},\Omega)\delta_{{\bf k},{\bf k}^{\prime}}-\Delta \mathbf{G}({\bf k};{\bf q},\Omega):\mathbf{K}({\bf k},{\bf k}^{\prime};{\bf q},\Omega)
\end{eqnarray}
and notations
\bea\label{parameters}
\mathbf{P}({\bf k};{\bf q}) & \equiv &
\frac{1}{2\imath\rho}\left(\mathbf{I}\otimes\mathbf{L}(\mathbf{k}^{-})-\mathbf{L}(\mathbf{k}^{+})\otimes\mathbf{I}\right)\\
\mathbf{\Delta <G>}({\bf k};{\bf q},\Omega) &\equiv &\frac{1}{2\imath\rho}\left(\mathbf{I}\otimes<\mathbf{G}>^{-}(\mathbf{k}^{-},\omega^{-})-<\mathbf{G}>^{+}(\mathbf{k}^{+},\omega^{+})\otimes\mathbf{I}\right) \nonumber \\
\mathbf{\Delta\Sigma}({\bf k};{\bf q},\Omega) &\equiv &
\frac{1}{2\imath\rho}\left(\mathbf{I}\otimes\mathbf{\Sigma}^{-}(\mathbf{k}^{-},\omega^{-})-\mathbf{\Sigma}^{+}(\mathbf{k}^{+},\omega^{+})\otimes\mathbf{I}\right) \nonumber\\
\mathbf{E}&\equiv &\mathbf{I}\otimes\mathbf{I}\quad \nonumber \\ \mathbf{L}(\mathbf{k}^{\pm})&\equiv & L_{ad}(\mathbf{k}^{\pm})=-c_{abcd}k^{\pm}_{b}k^{\pm}_{c}\nonumber
\eea
}
\subsubsection{pre-WTI}
\label{sec:pre-WTI}
As is well-known (see, for example \cite{Barabanenkov1991,Barabanenkov1995}), a Ward-Takahashi identity (WTI), that mathematically relates $\mathbf{\Sigma}$ to $\mathbf{K}$ and physically reflects conservative laws, which are satisfied in a particular physical system, and therefore, restricts classes of relevant scattering potentials,  should be established in order to properly treat Eqn. (\ref{BSfinal}). To this end, a preliminary identity (``pre-WTI''), Eqn. (\ref{preWTI}) below, will be obtained.

We start with two equations for Green tensors at two different set of variables: $(\vec x_{1},\vec x^{\prime}_{1},\omega_{1})$ and $(\vec x_{2},\vec x^{\prime}_{2},\omega_{2})$. It reads as follows
\bea\label{greeneqI}
\rho\omega_{1}^{2} G_{a_1e_1}(\vec x_{1},\vec x^{\prime}_{1},\omega_{1})+c_{a_1b_1c_1d_1}\frac{\partial^2}{\partial x_{b1}\partial x_{c1}}G_{d_1e_1}(\vec x_{1},\vec x^{\prime}_{1},\omega_{1}) &=& \\
&& \hspace{-6em} -V_{a_1d_1}(\vec x_{1},\omega_{1})G_{d_1e_1}(\vec x_{1},\vec x^{\prime}_{1},\omega_{1})-\delta_{a_1e_1}\delta(\vec x_{1}-\vec x^{\prime}_{1}) \nonumber \\
\label{greeneqII}
\rho\omega_{2}^{2} G_{a_2e_2}(\vec x_{2},\vec x^{\prime}_{2},\omega_{2})+c_{a_2b_2c2d2}\frac{\partial^2}{\partial x_{b2}\partial x_{c2}}G_{d_2e_2}(\vec x_{2},\vec x^{\prime}_{2},\omega_{2}) &=&  \\
&& \hspace{-6em} -V_{a_2d_2}(\vec x_{2},\omega_{2})G_{d_2e_2}(\vec x_{2},\vec x^{\prime}_{2},\omega_{2})-\delta_{a_2e_2}\delta(\vec x_{2}-\vec x^{\prime}_{2}). \nonumber
\eea
Then, we take the inverse Fourier transform from both expressions according to the rule
\begin{eqnarray}\label{ift}
\mathbf{F}(\vec k,\vec k^{\prime};\omega)=\int\int d\vec x d\vec x^{\prime} e^{-\imath \vec k\vec x}\mathbf{F}(\vec x,\vec x^{\prime};\omega) e^{\imath \vec k^{\prime}\vec x^{\prime}}
\end{eqnarray}
and substitute {the advanced} Green tensor Fourier transform at the terms, where operators act on it, as follows
\begin{eqnarray}\label{dft}
\mathbf{G}(\vec x,\vec x^{\prime};\omega)=\int\limits_{\vec k} \int\limits_{\vec k^{\prime}}e^{\imath \vec k\vec x} \mathbf{G}(\vec k,\vec k^{\prime};\omega)e^{-\imath \vec k^{\prime}\vec x^{\prime}}
\end{eqnarray}
It yields
\bea\label{greeneqIift}
&& \hspace{-4em}\rho\omega_{1}^{2} G_{a_1e_1}(\vec k_{1},\vec k^{\prime}_{1},\omega_{1})-c_{a_1b_1c_1d_1}k_{b_1}k_{c_1}G_{d_1e_1}(\vec k_{1},\vec k^{\prime}_{1},\omega_{1}) \\ \vspace{1em} \nonumber \\
&=&-\int\limits_{\vec p_{1}}\int d\vec x_{1}e^{-\imath \vec k_{1}\vec x_{1}}
V_{a1d1}(\vec x_{1},\omega_{1})e^{\imath \vec p_{1}\vec x_{1}}\int\limits_{\vec p^{\prime}_{1}} G_{d_1e_1}(\vec p_{1},\vec p^{\prime}_{1};\omega_{1})\int d\vec x^{\prime}_{1}e^{-\imath (\vec p^{\prime}_{1}-\vec k^{\prime}_{1})\vec x^{\prime}_{1}}-\delta_{a1e1}\delta_{\vec k_{1},\vec k^{\prime}_{1}} \nonumber\\
&=&-\sum\limits_{n=1}^{N}\int\limits_{\vec p_{1}}\int d\vec x_{1}e^{-\imath \vec k_{1}\vec x_{1}}V^{n}_{a1d1}(\vec x_{1},\omega_{1})e^{\imath \vec p_{1}\vec x_{1}}G_{d_1e_1}(\vec p_{1},\vec k^{\prime}_{1};\omega_{1})-\delta_{a_1e_1}\delta_{\vec k_{1},\vec k^{\prime}_{1}}\nonumber\\
&=& g(\omega_{1})\sum\limits_{n=1}^{N}\int\limits_{\vec p_{1}}\frac{\mu^2 b^2}{M} \mM_{a_1h_1}^{n}k_{h_1}p_{g_1} \mM_{g_1d_1}^{n}e^{i \left(\vec p_{1}-\vec k_{1} \right)\cdot \vec X^{n}_0}G_{d_1e_1}(\vec p_{1},\vec k^{\prime}_{1};\omega_{1})-\delta_{a_1e_1}\delta_{\vec k_{1},\vec k^{\prime}_{1}}\nonumber
\eea
{with $g(\omega_{1})=\frac{1}{\omega_{1}^{2} {+} i \omega_{1} (B/M) -\omega^2_0 }$.
To underline the fact that frequency dependent part $g(\omega_{1})$ of ${\mathcal A}^{n}(\omega_{1})$ does not depend on $n$, we  take it  out of the summation over $n$.
Acting similarly for the second set of variables one can come to the following system of equations
\bea
-(G^{0})^{-1}_{a_1d_1}(\vec k_{1},\omega_{1})G_{d_1e_1}(\vec k_{1},\vec k^{\prime}_{1},\omega_{1})+\delta_{a_1e_1}\delta_{\vec k_{1},\vec k^{\prime}_{1}}&=&  \\
&& \hspace{-8em}
g(\omega_{1})\sum\limits_{n=1}^{N}\int\limits_{\vec p_{1}}\frac{\mu^2 b^2}{M} M_{a_1h_1}^{n}k_{h_1}p_{g_1} M_{g_1d_1}^{n}e^{i \left(\vec p_{1}-\vec k_{1} \right)\cdot \vec X^{n}_0}G_{d_1e_1}(\vec p_{1},\vec k^{\prime}_{1};\omega_{1}) \nonumber \\
-(G^{0})^{-1}_{a_2d_2}(\vec k_{2},\omega_{2})G_{d_2e_2}(\vec k_{2},\vec k^{\prime}_{2},\omega_{2})+\delta_{a_2e_2}\delta_{\vec k_{2},\vec k^{\prime}_{2}}&=&  \nonumber \\
&& \hspace{-8em}
g(\omega_{2})\sum\limits_{n=1}^{N}\int\limits_{\vec p_{2}}\frac{\mu^2 b^2}{M} M_{a_2h_2}^{n}k_{h_2}p_{g_2} M_{g_2d_2}^{n}e^{i \left(\vec p_{2}-\vec k_{2} \right)\cdot \vec X^{n}_0}G_{d_2e_2}(\vec p_{2},\vec k^{\prime}_{2};\omega_{2})\nonumber
\label{greeneqsysift}
\eea
with
\bea
(G^{0})^{-1}_{ad}(\vec k,\omega)=-\left(\rho\omega^{2}\delta_{ad}-c_{abcd}k_{b}k_{c}\right).
 \label{g0ft}
\eea
{So far, we have played only with advanced Green tensors, whereas in order to construct intensity we need to involve the equation for the retarded Green tensor. To achieve this we make a complex conjugate of the second equation of the system, which eventually goes to the form:
\bea
-(G^{0})^{-1}_{a_1d_1}(\vec k_{1},\omega_{1})G_{d_1e_1}(\vec k_{1},\vec k^{\prime}_{1},\omega_{1})+\delta_{a_1e_1}\delta_{\vec k_{1},\vec k^{\prime}_{1}}&=& \\
&& \hspace{-10em} g(\omega_{1})\sum\limits_{n=1}^{N}\int\limits_{\vec p_{1}}\frac{\mu^2 b^2}{M} M_{a_1h_1}^{n}k_{h_1}p_{g_1} M_{g_1d_1}^{n}e^{i \left(\vec p_{1}-\vec k_{1} \right)\cdot \vec X^{n}_0}G_{d_1e_1}(\vec p_{1},\vec k^{\prime}_{1};\omega_{1})\nonumber\\
-(G^{0*})^{-1}_{a_2d_2}(\vec k_{2},\omega_{2})G^{*}_{d_2e_2}(\vec k_{2},\vec k^{\prime}_{2},\omega_{2})+\delta_{a_2e_2}\delta_{\vec k_{2},\vec k^{\prime}_{2}}&=&\nonumber\\
&& \hspace{-10em} g^{*}(\omega_{2})\sum\limits_{n=1}^{N}\int\limits_{\vec p_{2}}\frac{\mu^2 b^2}{M} M_{a_2h_2}^{n}k_{h_2}p_{g_2} M_{g_2d_2}^{n}e^{-i \left(\vec p_{2}-\vec k_{2} \right)\cdot \vec X^{n}_0}G^{*}_{d_2e_2}(\vec p_{2},\vec k^{\prime}_{2};\omega_{2})\nonumber
\label{greeneqsysiftcc}
\eea
Now we act on the first and second equations of the system from the right by $g^{*}(\omega_{2})(G)^{-1}_{e1f1}(\vec k^{\prime}_{1},\vec k^{\prime\prime}_{1};\omega_{1})$ and $g(\omega_{1})(G^{*})^{-1}_{e2f2}(k^{\prime}_{2},\vec k^{\prime\prime}_{2};\omega_{2})$, respectively.
It leads to the following system
\bea
-(G^{0})^{-1}_{a_1f_1}(\vec k_{1},\omega_{1})\delta_{\vec k_{1},\vec k^{\prime\prime}_{1}}g^{*}(\omega_{2})+g^{*}(\omega_{2})(G)^{-1}_{a_1f_1}(\vec k_{1},\vec k^{\prime\prime}_{1};\omega_{1})&=&\\
&& \hspace{-8em} g(\omega_{1})g^{*}(\omega_{2})\sum\limits_{n=1}^{N}\frac{\mu^2 b^2}{M} M_{a_1h_1}k_{h_1}k^{\prime\prime}_{g_1} M_{g_1f_1}e^{i \left(\vec k^{\prime\prime}_{1}-\vec k_{1} \right)\cdot \vec X^{n}_0}\nonumber\\
-(G^{0*})^{-1}_{a_2f_2}(\vec k_{2},\omega_{2})\delta_{\vec k_{2},\vec k^{\prime\prime}_{2}}g(\omega_{1})+g(\omega_{1})(G^{*})^{-1}_{a_2f_2}(\vec k_{2},\vec k^{\prime\prime}_{2};\omega_{2})&=&\nonumber\\
&& \hspace{-8em} g(\omega_{1})g^{*}(\omega_{2})\sum\limits_{n=1}^{N}\frac{\mu^2 b^2}{M} M_{a_2h_2}k_{h_2}k^{\prime\prime}_{g_2} M_{g_2f_2}e^{-i \left(\vec k^{\prime\prime}_{2}-\vec k_{2} \right)\cdot \vec X^{n}_0}\nonumber
\label{greeneqsysift}
\eea
{The next step consists in subtraction {at $B=0$ (so that $g(\omega)=g^{*}(\omega)$)} of the second equation from the first one and {evaluating at} $a_1=f_2=a$, $f_1=a_2=f$;
$\vec k_{2}\rightarrow\vec k^{\prime\prime}_{1}$, $\vec k^{\prime\prime}_{2}\rightarrow\vec k_{1}$.  Otherwise, it is not possible to eliminate the remaining parts of the potentials in both equations that are subjected to the subtraction from each other, since the parts imply not only summation over defects but also contain components of the second rank tensor, i. e. to achieve identity of those parts between each other, the components must be also identical.
{As a result, we have
\bea\label{preWTIft}
\lim\limits_{{\substack{\vec k_{2}\rightarrow\vec k^{\prime\prime}_{1}\\ \vec k^{\prime\prime}_{2}\rightarrow \vec k_{1}}}}
\left(-(G^{0})^{-1}_{af}(\vec k_{1},\omega_{1})\delta_{\vec k_{1},\vec k^{\prime\prime}_{1}}g^{*}(\omega_{2})+g^{*}(\omega_{2})(G)^{-1}_{af}(\vec k_{1},\vec k^{\prime\prime}_{1};\omega_{1})+\right.&&\nonumber\\ \left.(G^{*0})^{-1}_{fa}(\vec k_{2},\omega_{2})\delta_{\vec k_{2},\vec k^{\prime\prime}_{2}}g(\omega_{1})-g(\omega_{1})(G^{*})^{-1}_{fa}(\vec k_{2},\vec k^{\prime\prime}_{2};\omega_{2})\right)&\equiv&0\nonumber
\eea
By acting on the obtained identity with $\lim\limits_{{\substack{\vec k_{2}\rightarrow\vec k^{\prime\prime}_{1}\\ \vec k^{\prime\prime}_{2}\rightarrow \vec k_{1}}}} G(\vec k^{\prime\prime}_{1},\vec k^{\prime\prime\prime}_{1} ;\omega_{1})_{fg}G^{*}(\vec k^{\prime\prime}_{2},\vec k^{\prime\prime\prime}_{2};\omega_{2})_{ah}$ from the right, we get
\bea\label{preWTIftprod}
\lim\limits_{{\substack{\vec k_{2}\rightarrow\vec k^{\prime\prime}_{1}\\ \vec k^{\prime\prime}_{2}\rightarrow \vec k_{1}}}}
\left(-G^{*}(\vec k^{\prime\prime\prime}_{2},\vec k^{\prime\prime}_{2} ;\omega_{2})_{ha}(G^{0})^{-1}_{af}(\vec k_{1},\omega_{1})G(\vec k_{1},\vec k^{\prime\prime\prime}_{1} ;\omega_{1})_{fg}g^{*}(\omega_{2})+g^{*}(\omega_{2})\delta_{\vec k_{1},\vec k^{\prime\prime\prime}_{1}}G^{*}(\vec k^{\prime\prime}_{2},\vec k^{\prime\prime\prime}_{2};\omega_{2})_{gh}\right.&&\nonumber\\ \left.+G(\vec k^{\prime\prime\prime}_{1},\vec k^{\prime\prime}_{1};\omega_{1})_{gf}(G^{*0})^{-1}_{fa}(\vec k_{2},\omega_{2})G^{*}(\vec k_{2},\vec k^{\prime\prime\prime}_{2} ;\omega_{2})_{ah}g(\omega_{1})-g(\omega_{1})\delta_{\vec k_{2},\vec k^{\prime\prime\prime}_{2}}G(\vec k^{\prime\prime}_{1},\vec k^{\prime\prime\prime}_{1} ;\omega_{1})_{hg}\right)&\equiv&0\nonumber
\eea
After making averaging and introducing notations :
\bea\label{notat}
\vec k_{1}=\vec k^{+}\quad\vec k^{\prime\prime}_{2}=\vec k^{-}\quad\vec k^{\prime\prime}_{1}=\vec k^{\prime\prime+}\quad\vec k_{2}=\vec k^{\prime\prime-}\quad\vec k^{\prime\prime\prime}_{1}=\vec k^{\prime\prime\prime+}\quad \vec k^{\prime\prime\prime}_{2}=\vec k^{\prime\prime\prime-}\\
\omega_{1}=\omega_{+}\quad\omega_{2}=\omega_{-}\quad G=G^{+}\quad  G^{*}=G^{-}\quad G^{0}=G^{0+}\quad G^{0*}=G^{0-}\nonumber
\eea
we come to the following result
\bea\label{preWTIftprod}
-{\int\limits_{\mathbf{k}}}<G^{-}(\vec k^{\prime\prime\prime-},\vec k^{-} ;\omega_{-})_{ha}(G^{0+})^{-1}_{af}(\vec k^{+},\omega_{+})G^{+}(\vec k^{+},\vec k^{\prime\prime\prime+} ;\omega_{+})_{fg}>g^{\ast}(\omega_{-})&&\\ +g^{\ast}(\omega_{-})<\delta_{\vec k,\vec k^{\prime\prime\prime}}G^{-}(\vec k^{-},\vec k^{\prime\prime\prime-};\omega_{-})_{gh}>&&\nonumber\\ +{\int\limits_{\mathbf{k}^{\prime\prime}}}<G^{-}(\vec k^{\prime\prime\prime-},\vec k^{\prime\prime-};\omega_{-})_{ha}(G^{0-})^{-1}_{af}(\vec k^{\prime\prime-},\omega_{-})G^{+}(\vec k^{\prime\prime+},\vec k^{\prime\prime\prime+} ;\omega_{+})_{fg}g(\omega_{+})>&& \nonumber\\
-g(\omega_{+})<\delta_{\vec k^{\prime\prime},\vec k^{\prime\prime\prime}}G^{+}(\vec k^{\prime\prime+},\vec k^{\prime\prime\prime+} ;\omega_{+})_{hg}> &\equiv&0\nonumber
\eea
If we use Eqn.(\ref{intensity}) in Eqn.(\ref{preWTIftprod}) and recall that
\begin{eqnarray}\label{intensityft}
(G^{0\pm})^{-1}_{af}(\vec k^{\pm},\omega_{\pm})&=&(G^{0\pm})^{-1}_{af}({\bf k};{\bf q},\Omega) \\
<G^{\pm}(\vec k^{\prime\prime\prime\pm},\vec k^{\prime\prime\prime\pm} ;\omega_{\pm})_{hg}>&=&<G>^{\pm}(\vec k^{\prime\prime\prime};{\bf q},\Omega)_{hg}\nonumber
\end{eqnarray}
then, the pre-WTI reads as follows
\bea\label{preWTI}
{\int\limits_{\mathbf{k}}}\left((G^{0-})^{-1}_{fa}({\bf k};{\bf q},\Omega)g(\omega_{+})-(G^{0+})^{-1}_{fa}({\bf k};{\bf q},\Omega)g^{\ast}(\omega_{-})\right)\Phi_{fa,gh}({\bf k},{\bf k}^{\prime\prime\prime};{\bf q},\Omega)&&\\ +g^{\ast}(\omega_{-})<G>^{-}(\vec k^{\prime\prime\prime};{\bf q},\Omega)_{gh}-g(\omega_{+})<G>^{+}(\vec k^{\prime\prime\prime};{\bf q},\Omega)_{gh}&\equiv&0\nonumber
\eea
{This is a relation between the averaged Green's function, and its two-point correlations.}}
\subsubsection{WTI}
{{We now turn the relation between averages obtained at the end of the last subsection into a relation between their "irreducible" parts.} In this subsection, based on Eqn. (\ref{preWTI}), we establish an explicit relationship between {the irreducible vertex} $\mathbf{K}$ and {the mass operator} $\mathbf{\Sigma}$ which is the essence of the WTI. Indeed, by acting with $\Phi^{-1}_{gh,bc}({\bf k}^{\prime\prime\prime},{\bf k}^{\prime\prime\prime\prime};{\bf q},\Omega)$ on Eqn. (\ref{preWTI}) from the right , we obtain
\bea\label{preWTIftprodIII}
\left((G^{0-})^{-1}_{bc}({\bf k}^{\prime\prime\prime\prime};{\bf q},\Omega)g(\omega_{+})-(G^{0+})^{-1}_{bc}({\bf k}^{\prime\prime\prime\prime};{\bf q},\Omega)g^{\ast}(\omega_{-})\right)&&\\ +g^{\ast}(\omega_{-}){\int\limits_{\mathbf{k}^{\prime\prime\prime}}}<G>^{-}(\vec k^{\prime\prime\prime};{\bf q},\Omega)_{gh}\Phi^{-1}_{gh,bc}({\bf k}^{\prime\prime\prime},{\bf k}^{\prime\prime\prime\prime};{\bf q},\Omega)&&\nonumber\\-g(\omega_{+}){\int\limits_{\mathbf{k}^{\prime\prime\prime}}}<G>^{+}(\vec k^{\prime\prime\prime};{\bf q},\Omega)_{gh}\Phi^{-1}_{gh,bc}({\bf k}^{\prime\prime\prime},{\bf k}^{\prime\prime\prime\prime};{\bf q},\Omega)&\equiv&0\nonumber
\eea
According to Eqn. (\ref{Irred}), we have
\beq
\label{bsft}
\Phi^{-1}_{gh,bc}({\bf k}^{\prime\prime\prime},{\bf k}^{\prime\prime\prime\prime};{\bf q},\Omega)=(<G>^{+})^{-1}(\vec k^{\prime\prime\prime};{\bf q},\Omega)_{gb} (<G>^{-})^{-1}(\vec k^{\prime\prime\prime};{\bf q},\Omega)_{ch}\delta_{\vec k^{\prime\prime\prime},k^{\prime\prime\prime\prime}}-
K_{gh,bc}({\bf k}^{\prime\prime\prime},{\bf k}^{\prime\prime\prime\prime};{\bf q},\Omega)
\eeq
It gives
\bea\label{preWTIftprodIII}
\left((G^{0-})^{-1}_{bc}({\bf k}^{\prime\prime\prime\prime};{\bf q},\Omega)g(\omega_{+})-(G^{0+})^{-1}_{bc}({\bf k}^{\prime\prime\prime\prime};{\bf q},\Omega)g^{\ast}(\omega_{-})\right)&&\\ +g^{\ast}(\omega_{-})
\left((<G>^{+})^{-1}(\vec k^{\prime\prime\prime\prime};{\bf q},\Omega)_{bc}-
{\int\limits_{\mathbf{k}^{\prime\prime\prime}}}<G>^{-}(\vec k^{\prime\prime\prime};{\bf q},\Omega)_{gh}K_{gh,bc}({\bf k}^{\prime\prime\prime},{\bf k}^{\prime\prime\prime\prime};{\bf q},\Omega)\right)&&\nonumber\\
-g(\omega_{+})
\left((<G>^{-})^{-1}(\vec k^{\prime\prime\prime\prime};{\bf q},\Omega)_{bc}-
{\int\limits_{\mathbf{k}^{\prime\prime\prime}}}<G>^{+}(\vec k^{\prime\prime\prime};{\bf q},\Omega)_{gh}K_{gh,bc}({\bf k}^{\prime\prime\prime},{\bf k}^{\prime\prime\prime\prime};{\bf q},\Omega)\right) &\equiv&0\nonumber
\eea
Eventually, using (\ref{Dyson}) the WTI {(through which, the incoherent properties, captured by $K$, are related to the coherent properties, described by $\Sigma$ and $<G>$) reads}
\bea\label{WTIevent}
&&\left(\Sigma^{-}_{bc}({\bf k}^{\prime\prime\prime\prime};{\bf q},\Omega)g(\omega_{+})-\Sigma^{+}_{bc}({\bf k}^{\prime\prime\prime\prime};{\bf q},\Omega)g^{\ast}(\omega_{-})\right)\equiv\\
&& \hspace{4em} {\int\limits_{\mathbf{k}^{\prime\prime\prime}}}\left(g^{\ast}(\omega_{-})
<G>^{-}(\vec k^{\prime\prime\prime};{\bf q},\Omega)_{gh}-g(\omega_{+})
<G>^{+}(\vec k^{\prime\prime\prime};{\bf q},\Omega)_{gh}\right) K_{gh,bc}({\bf k}^{\prime\prime\prime},{\bf k}^{\prime\prime\prime\prime};{\bf q},\Omega)\nonumber
\eea
{It is worth mentioning that the WTI immediately reduces to the optical theorem when $B=0$, i.e. $g$ is real,  ${\bf q}, \Omega$ tend to zero, and the standard ISA expressions for  ${\bf \Sigma}$ and ${\bf K}$ tensors, {Eqn. (\ref{WTIisaexpressions}) below}, are taken (See \ref{OT}).Explicitly, the WTI reads in this case
\bea\label{WTIeventzero}
\left(\Sigma^{\ast}_{bc}({\bf k}^{\prime\prime\prime\prime})-\Sigma_{bc}({\bf k}^{\prime\prime\prime\prime})\right)\equiv {\int\limits_{\mathbf{k}^{\prime\prime\prime}}}\left(G^{0\ast}(\vec k^{\prime\prime\prime})_{gh}-G^{0}(\vec k^{\prime\prime\prime})_{gh}\right)
K_{gh,bc}({\bf k}^{\prime\prime\prime},{\bf k}^{\prime\prime\prime\prime})
\eea
with the following expressions, valid to leading order in $n$, the density of scatterers:
\bea
\label{WTIisaexpressions}
\Sigma_{bc}({\bf k}^{\prime\prime\prime\prime})=\Sigma_{bc}({\bf k}^{\prime\prime\prime\prime};{\bf 0},0)&=&n<t>_{bc}({\bf k}^{\prime\prime\prime\prime}) \nonumber \\
K_{gh,bc}({\bf k}^{\prime\prime\prime},{\bf k}^{\prime\prime\prime\prime};{\bf 0},0)=K_{gh,bc}({\bf k}^{\prime\prime\prime},{\bf k}^{\prime\prime\prime\prime})&=&n<t_{gb}({\bf k}^{\prime\prime\prime},{\bf k}^{\prime\prime\prime\prime})t^{*}_{ch}({\bf k}^{\prime\prime\prime\prime},{\bf k}^{\prime\prime\prime})>\nonumber\\
<G>(\vec k^{\prime\prime\prime};{\bf 0},0)_{hg}=<G>(\vec k^{\prime\prime\prime})_{hg}&\rightarrow& G^{0}(\vec k^{\prime\prime\prime})_{hg}
\eea

In terms of the general, i. e. symbolical, representation of the WTI there are two differences compared to a well-known tensorial version of the WTI for {electromagnetic} waves \cite{Barabanenkov1995}:
first, $g$ is a complex valued resonance like function; second, a tensor rank of the WTI is two rather than four as in the case of {electromagnetic} waves \cite{Barabanenkov1995}. {In our case, this is all we need to solve the problem at hand.}

We now change to a compact notation for the WTI by introducing
\bea\label{gnotation}
g(\omega_{+})=\frac{g(\omega_{+})+g^{\ast}(\omega_{-})}{2}+\frac{g(\omega_{+})-g^{\ast}(\omega_{-})}{2}\\
g^{\ast}(\omega_{-})=\frac{g(\omega_{+})+g^{\ast}(\omega_{-})}{2}-\frac{g(\omega_{+})-g^{\ast}(\omega_{-})}{2}\nonumber
\eea
and come to the WTI in the form
\bea\label{WTIcommonform}
\Sigma^{-}_{bc}({\bf k}^{\prime\prime\prime\prime};{\bf q},\Omega)-\Sigma^{+}_{bc}({\bf k}^{\prime\prime\prime\prime};{\bf q},\Omega)-{\int\limits_{\mathbf{k}^{\prime\prime\prime}}}\left(<G>^{-}(\vec k^{\prime\prime\prime};{\bf q},\Omega)_{gh}-<G>^{+}(\vec k^{\prime\prime\prime};{\bf q},\Omega)_{gh}\right)\\
\times K_{gh,bc}({\bf k}^{\prime\prime\prime},{\bf k}^{\prime\prime\prime\prime};{\bf q},\Omega)
\equiv\left(\frac{g^{\ast}(\omega_{-})-g(\omega_{+})}{g(\omega_{+})+g^{\ast}(\omega_{-})}\right)
\left(\Sigma^{-}_{bc}({\bf k}^{\prime\prime\prime\prime};{\bf q},\Omega)+\Sigma^{+}_{bc}({\bf k}^{\prime\prime\prime\prime};{\bf q},\Omega)\right.\nonumber\\ +\left.{\int\limits_{\mathbf{k}^{\prime\prime\prime}}}\left(<G>^{-}(\vec k^{\prime\prime\prime};{\bf q},\Omega)_{gh}+<G>^{+}(\vec k^{\prime\prime\prime};{\bf q},\Omega)_{gh}\right)
K_{gh,bc}({\bf k}^{\prime\prime\prime},{\bf k}^{\prime\prime\prime\prime};{\bf q},\Omega)\right)\nonumber
\eea
Eqn. (\ref{WTIcommonform}) can be rewritten in a more compact form, as
\beq\label{WLWTI}
\int\limits_{\mathbf{k}^{\prime\prime}}\mathbf{\overline{U}}(\mathbf{k}^{\prime\prime},\mathbf{k}^{\prime};\mathbf{q},\Omega)= \frac{i}{2}\mathbf{\overline{A}}(\mathbf{k}^{\prime};\mathbf{q},\Omega)\left(g(\omega_{+})-g^{\ast}(\omega_{-})\right)
\eeq
with
\begin{eqnarray}\label{WLUtensor}
\mathbf{\overline{U}}(\mathbf{k}^{\prime\prime},\mathbf{k}^{\prime};\mathbf{q},\Omega)&=&U_{aa,cd}({\bf k},{\bf k}^{\prime};{\bf q},\Omega)\\
\mathbf{\overline{A}}(\mathbf{k}^{\prime};\mathbf{q},\Omega)&=&A_{bb,cd}(\mathbf{k}^{\prime};\mathbf{q},\Omega)\nonumber\\
 \mathbf{A}(\mathbf{k}^{\prime};\mathbf{q},\Omega)&=&\frac{2}{g(\omega_{+})+g^{\ast}(\omega_{-})}
\left({\cal R}\mathbf{\Sigma}(\mathbf{k}^{\prime};\mathbf{q},\Omega)+\int\limits_{\mathbf{k}^{\prime\prime}}{\cal R}\mathbf{G}(\mathbf{k}^{\prime\prime};\mathbf{q},\Omega):\mathbf{K}({\bf k^{\prime\prime}},{\bf k^{\prime}};{\bf q},\Omega)\right) \, , \nonumber\\
{\cal R}\mathbf{\Sigma}(\mathbf{k}^{\prime};\mathbf{q},\Omega)&=&
\frac{1}{2\rho}\left(\mathbf{I}\otimes\mathbf{\Sigma}^{-}(\mathbf{k}^{\prime-},\omega^{-})+\mathbf{\Sigma}^{+}(\mathbf{k}^{\prime+},\omega^{+})\otimes\mathbf{I} \right) \, , \nonumber
\end{eqnarray}
{with tensor $U$ given by Eqn.(\ref{potBS})} and the operation ${\cal R}$, here defined for the self-energy tensor $\mathbf{\Sigma}$, acts in the same way on the Green tensor $\mathbf{G}$.

\subsection{Low frequency asymptotics and diffusion behavior}
\label{lowfreq}
The following relations for the self energy and for the Green function will prove useful:
\bea\label{SEDiff}
\mathbf{\Delta\Sigma}({\bf k};{\bf 0},0)
 & = & \mathbf{\Delta\Sigma({\bf k})}=\Delta\Sigma_{ab,cd}({\bf k})=\frac{1}{2\imath\rho}\left(\delta_{ac}\Sigma^{*}_{db}(\mathbf{k})-\Sigma_{ac}(\mathbf{k})\delta_{db}\right)\\
 & = & \frac{\delta_{ac}\delta_{db}}{2\imath\rho}\left(\Sigma^{*}(\mathbf{k})-\Sigma(\mathbf{k})\right) \nonumber\\
 & = & \delta_{ac}\delta_{db}\nonumber\\
 &   &\left(\frac{\rho c^{2}_{T}k^{2}}{4i\rho}\left(\left(1-\frac{\omega^{2}}{(K_{T}^{2})^{*}c^{2}_{T}}\right)-\left(1-\frac{\omega^{2}}{K_{T}^{2}c^{2}_{T}}\right)\right)+\right.\nonumber\\ &&\left.
 \frac{\rho c^{2}_{L}k^{2}}{4i\rho}\left(\left(1-\frac{\omega^{2}}{(K_{L}^{2})^{*}c^{2}_{L}}\right)-\left(1-\frac{\omega^{2}}{K_{L}^{2}c^{2}_{L}}\right)\right)\right) \nonumber\\
& = & -\frac{\delta_{ac}\delta_{db}\omega^{2} k^{2}}{2}\left(\frac{Im[K_{T}^{2}]}{K_{T}^{2}(K_{T}^{2})^{*}}+\frac{Im[K_{L}^{2}]}{K_{L}^{2}(K_{L}^{2})^{*}}\right) \, .\nonumber \\
\eea
\bea
\label{GFDiff}
\mathbf{\Delta G}({\bf k};{\bf 0},0)
 & = & \mathbf{\Delta G({\bf k})}=\Delta G_{ab,cd}({\bf k})=\frac{1}{2\imath\rho}\left(\delta_{ac}G^{*}_{db}(\mathbf{k})-G_{ac}(\mathbf{k})\delta_{db}\right)\\
& = &
\frac{\delta_{ac}}{2\imath\rho^{2}\omega^{2}}\left(\frac{(\delta_{db}-\hat k_d \hat k_b)}{\left\{\frac{k^2}{(K_{T}^{2})^{*}}-1\right\}}+\frac{\hat k_d \hat k_b}{\left\{\frac{k^2}{(K_{L}^{2})^{*}}-1\right\}}\right)-
\frac{\delta_{db}}{2\imath\rho^{2}\omega^{2}}\left(\frac{(\delta_{ac}-\hat k_a \hat k_c)}{\left\{\frac{k^2}{K_{T}^{2}}-1\right\}}+\frac{\hat k_a \hat k_c}{\left\{\frac{k^2}{K_{L}^{2}}-1\right\}}\right)\Rightarrow\nonumber\\
\Delta G_{aa,cd}({\bf k})
 & = &
 \frac{1}{2\imath\rho^{2}\omega^{2}}\left(\frac{(\delta_{dc}-\hat k_d \hat k_c)}{\left\{\frac{k^2}{(K_{T}^{2})^{*}}-1\right\}}-\frac{(\delta_{dc}-\hat k_d \hat k_c)}{\left\{\frac{k^2}{K_{T}^{2}}-1\right\}}+\frac{\hat k_d \hat k_c}{\left\{\frac{k^2}{(K_{L}^{2})^{*}}-1\right\}}-\frac{\hat k_d \hat k_c}{\left\{\frac{k^2}{K_{L}^{2}}-1\right\}}\right)\nonumber\\
& = &
 \frac{(\delta_{dc}-\hat k_d \hat k_c)}{2\imath\rho^{2}\omega^{2}}\left(\frac{(K_{T}^{2})^{*}}{k^{2}-(K_{T}^{2})^{*}}-\frac{K_{T}^{2}}{k^{2}-K_{T}^{2}}\right)+
 \frac{\hat k_d \hat k_c}{2\imath\rho^{2}\omega^{2}}\left(\frac{(K_{L}^{2})^{*}}{k^{2}-(K_{L}^{2})^{*}}-\frac{K_{L}^{2}}{k^{2}-K_{L}^{2}}\right)\nonumber\\
 & = & \frac{-(\delta_{dc}-\hat k_d \hat k_c)k^{2}}{\rho^{2}\omega^{2}}\frac{Im[K_{T}^{2}]}{\left(k^{2}-(K_{T}^{2})^{*}\right)\left(k^{2}-K_{T}^{2}\right)}+
 \frac{-\hat k_d \hat k_c k^{2}}{\rho^{2}\omega^{2}}\frac{Im[K_{L}^{2}]}{\left(k^{2}-(K_{L}^{2})^{*}\right)\left(k^{2}-K_{L}^{2}\right)} \nonumber \\
 & \approx & \frac{-\pi(\delta_{dc}-\hat k_d \hat k_c)k^{2}}{\rho^{2}\omega^{2}}\delta\left(k^{2}-Re[K_{T}^{2}]\right)+\frac{-\pi\hat k_d \hat k_c k^{2}}{\rho^{2}\omega^{2}}\delta\left(k^{2}-Re[K_{L}^{2}]\right)
 \label{deltaGdelta}
\eea
with
\beq
\Sigma=\rho\left(c_{T,L}^{2}-\frac{\omega^{2}}{K_{T,L}^{2}}\right)k^{2}\nonumber \, ,
\eeq
where {we introduced the shorthand notations for $<G^{+}>_{db}({\bf k},\omega) = G_{db}({\bf k})$ and $<G^{-}>_{db}({\bf k},\omega) = G^{*}_{db}({\bf k}))$ and in the same way for $\Sigma^{\pm}_{db}({\bf k},\omega)$. It should be emphasized that} the last approximate equality {in Eqn. (\ref{deltaGdelta})} holds in the limit $| Im[K_{T,L}^{2}] | \ll | k^{2}-Re[K_{T,L}^{2}] |$. {The meaning of this inequality in terms of the dislocation parameters is explored in  Section \ref{discone}.}

\subsection{Perturbation approach to BS eigenvalue problem}
In this section we follow the approach that was used in Refs.
 \cite{Stark1997,Barabanenkov1991,Barabanenkov1995,Berman2000} to study the diffusion of  electromagnetic and acoustic waves: The BS equation written in the form of Eqn. (\ref{BSfinal}), supplemented by {the relation between mass operator $\mathbf{\Sigma}$ and kernel $\mathbf{K}$ implemented by the WTI Eqn. (\ref{WLWTI})},} is solved for the intensity $\mathbf{\Phi}$, defined by Eqn. (\ref{intensity}), in the diffusive limit.  To this end, the BS equation Eqn. (\ref{BSfinal}) is written in operator form:
  \beq\label{BSH}
\int\limits_{\bf{k}^{\prime\prime}}\mathbf{H}({\bf k},{\bf k}^{\prime\prime};{\bf q},\Omega):\mathbf{\Phi}({\bf k}^{\prime\prime},{\bf k}^{\prime};{\bf q},\Omega)=\mathbf{\Delta G}({\bf k};{\bf q},\Omega)\delta_{{\bf k},{\bf k}^{\prime}} \, .
\eeq
with the operator $\mathbf{H}$ defined by
\beq\label{Hoperator}
\mathbf{H}({\bf k},{\bf k}^{\prime\prime};{\bf q},\Omega)\equiv \left[\imath\omega\Omega\mathbf{E}+\mathbf{P}({\bf k};{\bf q})\right]\delta_{\mathbf{k}\mathbf{k}^{\prime\prime}}+\mathbf{U}({\bf k},{\bf k}^{\prime\prime};{\bf q},\Omega) \, .
\eeq
It is easy to see, using the explicit form of $\mathbf{U}$, and the reciprocity of the tensor $\mathbf{K}$, that $\mathbf{H}$ has the following symmetry:
 \beq\label{symmetry}
H_{ab,cd}({\bf k},{\bf k}^{\prime\prime};{\bf q},\Omega)\Delta G_{cd,ef}({\bf k}^{\prime\prime};{\bf q},\Omega)=H_{ef,cd}({\bf k}^{\prime\prime},{\bf k};{\bf q},\Omega)\Delta G_{cd,ab}({\bf k};{\bf q},\Omega) \, .
\eeq
The solution of the BS equation will be found in terms of the eigenvalues and eigentensors of the operator $\mathbf{H}$.

The eigenvalue problem for $\mathbf{H}$ is set up as follows:
\beq
\label{Hom}
\int\limits_{\bf{k}^{\prime\prime}}H_{ab,cd}({\bf k},{\bf k}^{\prime\prime};{\bf q},\Omega)f^{\mathbf{r}n}_{cd}({\bf k}^{\prime\prime};{\bf q},\Omega)=
\lambda_{n}\left(\mathbf{q},\Omega\right)f^{\mathbf{r}n}_{ab}({\bf k};{\bf q},\Omega)
\eeq
where $\mathbf{f}^{\mathbf{r}n}({\bf k}^{\prime\prime};{\bf q},\Omega)$  (resp. $\mathbf{f}^{\mathbf{l}n}({\bf k}^{\prime\prime};{\bf q},\Omega)$) are right (resp. left) eigentensors and $\lambda_{n}\left(\mathbf{q},\Omega\right)$ the corresponding eigenvalues. Following \cite{Stark1997,Barabanenkov1991,Barabanenkov1995,Berman2000} we shall assume that the eigentensors in Eqn. (\ref{Hom}) satisfy completeness and orthogonality conditions:
\begin{eqnarray}\label{basis}
\int\limits_{\mathbf{k}}f^{\mathbf{r}m}_{ab}({\bf k};{\bf q},\Omega)f^{\mathbf{l}n}_{ab}({\bf k};{\bf q},\Omega)&=& \delta_{mn}\, ,\\
\sum\limits_{n}f^{\mathbf{r}n}_{ab}({\bf k};{\bf q},\Omega)f^{\mathbf{l}n}_{ef}({\bf k}^{\prime};{\bf q},\Omega)&=&\delta_{{\bf k}{\bf k}^{\prime}}\delta_{ae}\delta_{fb}\nonumber \, .
\end{eqnarray}
Furthermore, the symmetry restriction for the operator $\mathbf{H}$ from Eqn. (\ref{symmetry}) determines the relation between left and right eigentensors:
\beq
\label{lr}
f^{\mathbf{r}n}_{ab}({\bf k};{\bf q},\Omega)=\Delta G_{ab,cd}({\bf k};{\bf q},\Omega)
f^{\mathbf{l}n}_{cd}({\bf k};{\bf q},\Omega) \, .
\eeq

The eigentensor properties  Eqns. (\ref{basis}, \ref{lr}) allow for a representation of the solution $\mathbf{\Phi}$ of Eqn. (\ref{BSH})  as a series over the states $n$  \cite{Stark1997,Barabanenkov1991,Barabanenkov1995,Berman2000} :
\begin{eqnarray}\label{solution}
\Phi_{ab,cd}=\sum\limits_{n}\frac{f^{\mathbf{r}n}_{ab}({\bf k};{\bf q},\Omega)f^{\mathbf{r}n}_{cd}({\bf k}^{\prime};{\bf q},\Omega)}{\lambda_{n}\left(\mathbf{q},\Omega\right)}
\end{eqnarray}
The existence of a diffusion regime assumes that in the limit $\mathbf{q}\rightarrow0$, $\Omega\rightarrow0$ the intensity $\mathbf{\Phi}$ has a pole structure given by the lowest eigenvalue asymptotics $\lambda_{0}\left(\mathbf{q}\rightarrow 0,\Omega\rightarrow 0\right)\rightarrow 0$ that is separated from a regular part \cite{Barabanenkov1991,Barabanenkov1995,Berman2000}. Therefore, the whole problem is reduced to the determination of coefficients of a perturbative expansion for $\lambda_{0}\left(\mathbf{q},\Omega\right)$ to second order in $\mathbf{q}$ and first order in $\Omega$ around the point $\mathbf{q}=0$, $\Omega=0$. To do this, Eqn. (\ref{Hom}) has to be treated perturbatively, with the condition that Eqns. (\ref{WLWTI}, \ref{symmetry}) hold at every order of the perturbation in $\mathbf{q}$, and $\Omega$ \cite{Barabanenkov1991,Barabanenkov1995,Berman2000}.

{In order to solve} Eq. (\ref{Hom}), we write, in a small $\Omega$ and a small $\mathbf{q}$ approximation,
\begin{eqnarray}\label{Pseries}
\mathbf{H}({\bf k},{\bf k}^{\prime\prime};{\bf q},\Omega) & = &
\mathbf{H}({\bf k},{\bf k}^{\prime\prime};{\bf 0},0)+\mathbf{H}^{1\Omega}({\bf k},{\bf k}^{\prime\prime};{\bf 0},\Omega)\nonumber \\
& & +\mathbf{H}^{1\mathbf{q}}({\bf k},{\bf k}^{\prime\prime};{\bf q},0)+\mathbf{H}^{2\mathbf{q}}({\bf k},{\bf k}^{\prime\prime};{\bf q},0) + \dots \, , \nonumber \\
& & \nonumber \\
\label{Pseries2}
\mathbf{f}^{\mathbf{r}0}({\bf k}^{\prime\prime};{\bf q},\Omega) & = &
\mathbf{f}({\bf k}^{\prime\prime};{\bf 0},0)+\mathbf{f}^{1\Omega}({\bf k}^{\prime\prime};{\bf 0},\Omega) \\
& & +\mathbf{f}^{1\mathbf{q}}({\bf k}^{\prime\prime};{\bf q},0)+\mathbf{f}^{2\mathbf{q}}({\bf k}^{\prime\prime};{\bf q},0) + \dots \nonumber \\
& & \nonumber \\
\lambda_{0}\left(\mathbf{q},\Omega\right) & = &
\lambda^{1\Omega}\left(\mathbf{0},\Omega\right)+\lambda^{1\mathbf{q}}\left(\mathbf{q},0\right)+\lambda^{2\mathbf{q}}\left(\mathbf{q},0\right) + \dots \nonumber
\label{lambda}
\end{eqnarray}
Substitution of the above expansions into Eqn. (\ref{Hom}) leads to the following set of coupled integral equations:
\begin{eqnarray}\label{system0}
\int\limits_{\bf{k}^{\prime\prime}}H_{ab,cd}({\bf k},{\bf k}^{\prime\prime})f_{cd}({\bf k}^{\prime\prime})& = & 0 \\
\label{Omega}
\int\limits_{\bf{k}^{\prime\prime}}\left(H_{ab,cd}({\bf k},{\bf k}^{\prime\prime})f^{1\Omega}_{cd}({\bf k}^{\prime\prime})
+  H^{1\Omega}_{ab,cd}({\bf k},{\bf k}^{\prime\prime})f_{cd}({\bf k}^{\prime\prime})\right) & = &
\lambda^{1\Omega}f_{ab}({\bf k}) \\
\label{system1q}
\int\limits_{\bf{k}^{\prime\prime}}\left(H_{ab,cd}({\bf k},{\bf k}^{\prime\prime})f^{1\mathbf{q}}_{cd}({\bf k}^{\prime\prime}) +H^{1\mathbf{q}}_{ab,cd}({\bf k},{\bf k}^{\prime\prime})f_{cd}({\bf k}^{\prime\prime})\right) & = & 0 \\
\label{system2q}
\int\limits_{\bf{k}}{\cal{B}} P_{aa,cd}({\bf k};{\bf q})f^{1\mathbf{q}}_{cd}({\bf k}) & = &
\lambda^{2\mathbf{q}}
\end{eqnarray}
Within Eqns. (\ref{system0}-\ref{system2q}) we have used a shorthand notation for the quantities appearing in Eqn. (\ref{Pseries}), in which the arguments $\mathbf{q}$ and $\Omega$ are omitted. The superscript indicates the variable and the order of perturbation  for the operator $\mathbf{H}$, right {eigentensor} $\mathbf{f}^{\mathbf{r}0}$, and eigenvalue $\lambda_{0}$. The disappearance in Eqns. (\ref{system0}-\ref{system2q}) of contributions from some terms that appear in Eqn. (\ref{Pseries}) is due to the implementation of symmetry and conservation restrictions coming from Eqns. (\ref{WLWTI}) and (\ref{symmetry}). Importantly, the first-order-in-wavenumber contribution to the eigenvalue vanishes: $\lambda^{1\mathbf{q}} = 0$.} Without this result there would be no diffusion behavior. {This algebra follows closely Ref. \cite{Churochkin2017}.}

With the aid of Eqns. (\ref{WLWTI}) and (\ref{system0}), the eigentensor $\mathbf{f}^{\mathbf{r}0}$ at $\mathbf{q}=0$, $\Omega=0$ can be found at once:
\beq\label{zeroorder}
f_{cd}({\bf k}^{\prime\prime})={\cal{B}}{\Delta G_{cd,gg}({\bf k}^{\prime\prime})}
\eeq
with
\beq
{\cal{B}}=\frac{1}{\sqrt{\int\limits_{\bf{v}}{\Delta G_{cc,gg}({\bf v})}}} .
\label{eq:B}
\eeq
{Note that $\int\limits_{\bf{v}}{\Delta G_{cc,gg}({\bf v})}$ is negative, but only ${\cal{B}}^{2}$ appears in the expression for the diffusion constant.}
Integrating Eqn. (\ref{Omega}) over ${\bf k}$ along with the subsequent implementation of the WTI from Eqn. (\ref{WLWTI}) at the corresponding order, leads to the expression for the eigenvalue $\lambda^{1\Omega}$
\begin{eqnarray}\label{firstomega}
i\omega\Omega\left(1+a\right)=
\lambda^{1\Omega}
\end{eqnarray}
where we have introduced a parameter $a$ defined by
\begin{eqnarray}
\label{aparameter}
a=\frac{1}{\int\limits_{\bf{k}}f_{hh}({\bf k})}\int\limits_{\bf{k}^{\prime\prime}} \frac{\left(A_{aa,cd}(\mathbf{k}^{\prime\prime};\mathbf{0},\Omega)\left(g(\omega_{+})-g^{\ast}(\omega_{-})\right)\right)^{1\Omega}}{2\omega\Omega}f_{cd}({\bf k}^{\prime\prime}) \, .
\end{eqnarray}
This is an analogue of the well-known parameter that appears in the diffusion of light, which, being positive, renormalizes the phase velocity downwards {to a lower value for a transport velocity}
\cite{Barabanenkov1991,Barabanenkov1995,Tiggelen1993,Livdan1996}. To see that our $a$ is indeed positive, replace  Eqns. (\ref{WLUtensor}) and (\ref{zeroorder}) into Eqn. (\ref{aparameter}) to obtain
\bea
\label{aparaprox}
a  &=&
\frac{-1}{\rho^{2}\left(\omega_{0}^{2}-\omega^{2}\right)\int\limits_{\bf{v}}{\Delta G_{cc,hh}({\bf v})}}\int\limits_{\bf{k}}
Im[{\Sigma_{ab}(\mathbf{k})G_{ab}}({\bf k})]\\
&\approx&
\frac{Re[K_{T}^{2}]\left(\frac{c_{T}^{2}}{\omega^{2}}Re[K_{T}^{2}]-1\right)+Re[K_{L}^{2}]\left(\frac{c_{L}^{2}}{\omega^{2}}Re[K_{L}^{2}]-1\right)}
{\left(\frac{\omega_{0}^{2}}{\omega^{2}}-1\right)\left(Re[K_{T}^{2}]+Re[K_{L}^{2}]\right)}\nonumber
\eea
where the last approximate equality is obtained using expression Eqn. (\ref{FReg}) for the function $F$ defined by Eqn. (\ref{ApSolDerivative1}) that enters the integrand of Eqn.(\ref{aparaprox}) as a consequence of the calculation of $Im[{\Sigma_{ab}(\mathbf{k})G_{ab}}({\bf k})]$. {Clearly $a>0$ if $\omega<\omega_{0}$.}

\subsubsection{Explicit form of the diffusion constant}
Within the spectral approach that we are using, Eqns. (\ref{solution}-\ref{system2q}) lead to the following expression for the singular part of the intensity, $\mathbf{\Phi}^{sing}$:
\beq\label{sing}
\Phi^{sing}_{ab,cd}=\frac{f^{\mathbf{r0}}_{ab}({\bf k};{\bf q},\Omega)f^{\mathbf{r0}}_{cd}({\bf k}^{\prime};{\bf q},\Omega)}{\lambda^{1\Omega}+\lambda^{2\mathbf{q}}}=\frac{f^{\mathbf{r0}}_{ab}({\bf k};{\bf q},\Omega)f^{\mathbf{r0}}_{cd}({\bf k}^{\prime};{\bf q},\Omega)}{\frac{\lambda^{1\Omega}}{-i\Omega}\left(-i\Omega+\frac{-i\Omega\lambda^{2\mathbf{q}}}{\lambda^{1\Omega}q^{2}}q^{2}\right)} \, .
\eeq
Then, with the assistance of Eqns. (\ref{firstomega},\ref{sing}) the diffusion constant can be identified as
\begin{eqnarray}\label{Dconstant}
D& \equiv & -\frac{i\Omega\lambda^{2\mathbf{q}}}{q^2\lambda^{1\Omega}}=-\frac{{\cal{B}}}{q^{2}\omega\left(1+a\right)}
\int\limits_{\bf{k}}P_{aa,cd}({\bf k};{\bf q})f^{1\mathbf{q}}_{cd}({\bf k}) \\
 &=& \frac{{\cal{B}}^{2}}{q^{2}\omega\left(1+a\right)}
\int\limits_{\bf{k}}P_{aa,cd}({\bf k};{\bf q})
\left(\int\limits_{{\bf k}_{2}}\Phi_{cd,gh}({\bf k},{\bf k}_{2})P_{gh,ee}(\mathbf{q};{\bf k}_{2})-\Delta G_{cd,ee}^{1\mathbf{q}}({\bf k})\right) \label{Dconstant2} \\
&\equiv & D^{\mathcal{R}} + D_{\Delta G^{1\mathbf{q}}}
\label{Dconstant3}
\end{eqnarray}
with ${\cal{B}}$ given by Eqn. (\ref{eq:B}). To obtain Eqn. (\ref{Dconstant2}), in which the diffusion constant is written as the sum of two terms, we have substituted  the values for $\lambda^{2\mathbf{q}}$, $\lambda^{1\Omega}$ given by Eqns. (\ref{system2q}) and (\ref{firstomega}). The last one ensued from the form of $f^{1\mathbf{q}}({\bf k})$ ({See \ref{solu}}).
Thus, the expression for the diffusion constant in Eqn. (\ref{Dconstant}) is the sum of two contributions, as defined in (\ref{Dconstant3}).

While the second term in Eqn. (\ref{Dconstant3}), $D_{\Delta G^{1\mathbf{q}}}$, can be calculated straightforwardly (See \ref{esti}),
 the calculation of the  first term, $D^{\mathcal{R}}$, is more laborious. {Indeed, it depends on the unknown tensor function $\mathbf{\Phi}$. However, it is not the complete tensor function that is needed, but an integrated form  over one of its variables that, as we now show, can be expressed as a function of the mass operator tensor $\mathbf{\Sigma}$ and the kernel tensor $\mathbf{K}$ using the BS equation and the WTI.} In our case, in order to {do this}  we apply, inspired by the treatment of light diffusion  \cite{Barabanenkov1992}, the method that uses an auxiliary tensor function $\mathbf{\Psi}_{,s}({\bf k})$ defined by the relation
\beq\label{DconstantAUX}
\mathbf{\Psi}_{,s}({\bf k})q_{s} \equiv \Psi_{cd,s}({\bf k})q_{s}\equiv
\int\limits_{{\bf k}^{\prime}}\Phi_{cd,gh}({\bf k},{\bf k}^{\prime})P_{gh,ee}(\mathbf{q};{\bf k}^{\prime})=
-\int\limits_{{\bf k}^{\prime}}\Phi_{cd,gh}({\bf k},{\bf k}^{\prime})\frac{1}{2i\rho}\frac{\partial L_{gh}({\bf k}^{\prime})}{\partial k^{\prime}_{s}}q_{s} \, .
\eeq
Then, from Eqns. (\ref{BSfinal}-\ref{potBS}) the following expression for $\mathbf{\Psi}_{,s}({\bf k})$
immediately follows:
\bea\label{DconstantAUXEQ}
P_{ab,cd}({\bf p})\Psi_{cd,s}({\bf p})+\Delta\Sigma_{ab,c_1d_1}({\bf p})\Psi_{c_1d_1,s}({\bf p})
 \hspace{3em} &&\\
-\int\limits_{\bf{p}^{\prime\prime}}
\Delta G_{ab,c_2d_2}({\bf p})K_{c_2d_2,c_1d_1}({\bf p},{\bf p}^{\prime\prime})\Psi_{c_1d_1,s}({\bf p}^{\prime\prime})&=&
-\Delta G_{ab,gh}({\bf p})\frac{1}{2i\rho}\frac{\partial L_{gh}({\bf p})}{\partial p_{s}}\nonumber
\eea
Eqn. (\ref{DconstantAUXEQ}) can be substantially simplified if we recall the explicit form of the free medium Green's function, Eqn.(\ref{g0Sol}), along with relations from Eqns. (\ref{Dyson}, \ref{parameters}) \cite{Maurel2004a}. Then,
\bea\label{greenexpl}
\Delta G_{ab,c_2d_2}({\bf p})&=&\frac{-1}{2i\rho}\left(G^{-1*}_{ac_3}({\bf p})\delta_{d_3 b}-\delta_{a c_3}G^{-1}_{d_3 b}({\bf p})\right)G_{d_2 d_3}({\bf p})G_{c_3 c_2}^{*}({\bf p})\\
&=&\left(\Delta\Sigma_{ab,c_3 d_3}({\bf p})+P_{ab,c_3 d_3}({\bf p})\right)G_{d_2 d_3}({\bf p})G_{c_3 c_2}^{*}({\bf p})\nonumber
\eea
As a next step, we define a tensor angular entity $\mathbf{\Upsilon}$ that {is analogous to the coefficient relating transport mean free path and  extinction length in the diffusion of electromagnetic waves \cite{Sheng1990,Tiggelen2000}:}
\beq\label{angular}
\Psi_{cd,s}({\bf p})q_{s}=G^{*}({\bf p})_{cg}G_{hd}({\bf p})\Upsilon_{gh}({\bf p},{\bf q}) \, ,
\eeq
and an integral equation for $\mathbf{\Upsilon}$ follows straightforwardly from Eqns. (\ref{DconstantAUXEQ}-\ref{angular}):
\beq\label{angulareq}
\Upsilon_{gh}({\bf p},{\bf q})-
\int\limits_{\bf{p}^{\prime\prime}}
K_{gh,c_1 d_1}({\bf p},{\bf p}^{\prime\prime})G_{c_1 g_1}^{*}({\bf p}^{\prime\prime})G_{h_1 d_1}({\bf p}^{\prime\prime})\Upsilon_{g_1 h_1}({\bf p}^{\prime\prime},{\bf q})=P_{gh,ee}({\bf p};{\bf q}) \, .
\eeq
Then, using Eqns. (\ref{DconstantAUX}), (\ref{angular})
,  $D^{\mathcal{R}}$ from Eqn. (\ref{Dconstant3}) can be written as
\beq\label{difften}
D^{\mathcal{R}} =
\frac{{\cal{B}}^{2}}{q^{2}\omega\left(1+a\right)}
\int\limits_{\bf{k}}P_{aa,cd}({\bf k};{\bf q})G_{cg}^{*}({\bf k})
G_{hd}({\bf k})\Upsilon_{gh}({\bf k},{\bf q})
\eeq
Furthermore, guided by the definition from Eqn. (\ref{angular}) we can make a conjecture that tensor $\mathbf{\Upsilon}({\bf p},{\bf q})$ should be linear in ${\bf q}$. Keeping in mind this property of $\mathbf{\Upsilon}({\bf p},{\bf q})$ we seek for the corresponding solution in the form
 \beq\label{angulsol}
\Upsilon_{gh}({\bf p},{\bf q})=\alpha P_{gh,ff}({\bf p};{\bf q})
\eeq
 Then,  by multiplying Eqn. (\ref{angulareq}) with  $P_{e_1 e_1,g_1 h_1}({\bf p};{\bf q})G_{g_1 g}({\bf p})G_{hh_1}^{*}({\bf p})$ from the left and integrating over the ${\bf p}$  we remain with the relation
 \beq\label{angularproof}
\alpha^{-1}=1-
\frac{\int\limits_{\bf{p}}\int\limits_{\bf{p}^{\prime\prime}}P_{ff,ab}({\bf p};{\bf q})G_{a g}({\bf p})G_{hb}^{*}({\bf p})
K_{gh,cd}({\bf p},{\bf p}^{\prime\prime})G_{ca}^{*}({\bf p}^{\prime\prime})G_{bd}({\bf p}^{\prime\prime})P_{ab,ee}({\bf p}^{\prime\prime},{\bf q})}{\int\limits_{\bf{k}}P_{e_1 e_1,g_1 h_1}({\bf k};{\bf q})G_{g_1 g}({\bf k})G_{hh_1}^{*}({\bf k})P_{gh,ee}({\bf k};{\bf q})}
\eeq
{where the ratio of two integrals is the analog of the $\langle \cos \theta \rangle$ term in the diffusion of electromagnetic waves \cite{Tiggelen2000}}. As a consequence, $D^{\mathcal{R}}$ can be represented as
\beq\label{diffin}
D^{\mathcal{R}} =
\frac{{\cal{B}}^{2}}{q^{2}\omega\left(1+a\right)}
\int\limits_{\bf{k}}\alpha P_{aa,cd}({\bf k};{\bf q})G_{cg}^{*}({\bf k})G_{hd}({\bf k})P_{gh,a_1 a_1}({\bf k};{\bf q})
\eeq
It must be noted here that $\alpha$ included in the general expression for the diffusion constant from  Eqn. (\ref{diffin}) can be evaluated explicitly using the
symmetry properties of the Green tensor, tensor $\mathbf{P}$, and the kernel $\mathbf{K}$ from Eqns. (\ref{green}), (\ref{parameters}),and (\ref{WTIisaexpressions}), respectively.
Indeed, those equations support the validity of the following relations:
\bea\label{WTIzeroproof}
K_{gh,c_1 d_1}({\bf p},-{\bf p}^{\prime\prime})G_{c_1 g_1}^{*}(-{\bf p}^{\prime\prime})G_{h_1 d_1}(-{\bf p}^{\prime\prime})P_{g_1 h_1,ee}(-{\bf p}^{\prime\prime},{\bf q})&=&  \\
&& \hspace{-8em}-K_{gh,c_1 d_1}({\bf p},{\bf p}^{\prime\prime})G_{c_1 g_1}^{*}({\bf p}^{\prime\prime})G_{h_1 d_1}({\bf p}^{\prime\prime})P_{g_1 h_1,ee}({\bf p}^{\prime\prime},{\bf q}) \nonumber
\eea
Hence
\beq
 \int\limits_{{\bf p}^{\prime\prime}}K_{gh,c_1 d_1}({\bf p},{\bf p}^{\prime\prime})G_{c_1 g_1}^{*}({\bf p}^{\prime\prime})G_{h_1 d_1}({\bf p}^{\prime\prime})P_{g_1 h_1,ee}({\bf p}^{\prime\prime},{\bf q})=0
 \eeq
 and consequently
 \beq
 \alpha= 1 \, .
\eeq
Therefore, $D^{\mathcal{R}}$ from the Eqn.(\ref{diffin}) can be brought into the form
\beq\label{DCFin}
D^{\mathcal{R}} =
\frac{{\cal{B}}^{2}}{q^{2}\omega\left(1+a\right)}
\int\limits_{\bf{k}}P_{aa,cd}({\bf k};{\bf q})G_{cg}^{*}({\bf k})G_{hd}({\bf k})P_{gh,a_1 a_1}({\bf k};{\bf q})
\eeq
Finally, using approximation from Eqn. (\ref{FReg}) for Eqn. (\ref{ApSolDerivative}) in Eqn. (\ref{DCFin}) we can write
\beq
\label{DCFinExpl0}
D^{\mathcal{R}} =\frac{-{\cal{B}}^{2}}{8\rho^{2}\omega^{5}\left(1+a\right)}
\left(c_{L}^{4}\left(\frac{Re[K_{L}^{2}]^{3}}{Im[K_{L}^{2}]}+Re[K_{L}^{2}]Im[K_{L}^{2}]\right)+c_{T}^{4}\left(\frac{Re[K_{T}^{2}]^{3}}{Im[K_{T}^{2}]}+Re[K_{T}^{2}]Im[K_{T}^{2}]\right)\right) \, .
\eeq
Then, the total diffusion constant reads as
\bea
\label{DCFintot}
D&=&D^{\mathcal{R}}+D_{\Delta G^{1\mathbf{q}}} \nonumber \\
&=&\frac{-{\cal{B}}^{2}}{8\rho^{2}\omega^{5}\left(1+a\right)}
\left(c_{L}^{4}\left(\frac{Re[K_{L}^{2}]^{3}}{Im[K_{L}^{2}]}+Re[K_{L}^{2}]Im[K_{L}^{2}]\right)+c_{T}^{4}\left(\frac{Re[K_{T}^{2}]^{3}}{Im[K_{T}^{2}]}+Re[K_{T}^{2}]Im[K_{T}^{2}]\right)\right)\nonumber\\
&& \hspace{4em}+\frac{{\cal{B}}^{2}\left(c_{L}^{2}-c_{T}^{2}\right)\left(Im[K_{L}^{2}]-Im[K_{T}^{2}]\right))}{16\rho^{2} \omega^{3}\left(1+a\right)} \, , \nonumber
\eea
and the leading term in {the limit of small} $Im[K_{T,L}^{2}]$ {is}
\bea
\label{DCFintotlead}
D^{lead}\approx\frac{-{\cal{B}}^{2}}{8\rho^{2}\omega^{5}\left(1+a\right)}
\left(c_{L}^{4}\frac{Re[K_{L}^{2}]^{3}}{Im[K_{L}^{2}]}+c_{T}^{4}\frac{Re[K_{T}^{2}]^{3}}{Im[K_{T}^{2}]}\right)
\eea
Explicitly, from Eqn.(\ref{eq:B}), we have
\bea\label{Bexpl}
-{\cal{B}}^{2}=\frac{4\rho^{2}\omega^{2}}{\left(Re[K_{T}^{2}]+Re[K_{L}^{2}]\right)}
\eea
Then, using Eqns.(\ref{deltaGdelta}, \ref{eq:B}, \ref{aparaprox}, \ref{DCFintotlead})
\beq
\label{DCFintotleadexpl}
D^{lead}\approx
\frac{1}{\left(1+\frac{Re[K_{T}^{2}]\left(\frac{c_{T}^{2}}{\omega^{2}}Re[K_{T}^{2}]-1\right)+Re[K_{L}^{2}]\left(\frac{c_{L}^{2}}{\omega^{2}}Re[K_{L}^{2}]-1\right)}
{\left(\frac{\omega_{0}^{2}}{\omega^{2}}-1\right)\left(Re[K_{T}^{2}]+Re[K_{L}^{2}]\right)}\right)} \times
 \frac{\left(c_{L}^{4}\frac{Re[K_{L}^{2}]^{3}}{Im[K_{L}^{2}]}+c_{T}^{4}\frac{Re[K_{T}^{2}]^{3}}{Im[K_{T}^{2}]}\right)}{2\omega^{3}\left(Re[K_{T}^{2}]+Re[K_{L}^{2}]\right)}
\eeq
In the low-frequency limit it reads as
\bea\label{DCFintotleadexpllf}
D^{lead}_{\omega\rightarrow 0}\approx
\left(\frac{v_{T}^{2}c_{L}^{4}}{\left(v_{L}^{2}+v_{T}^{2}\right)v_{L}^{4}}\frac{v_{L}l_{L}}{2}+\frac{v_{L}^{2}c_{T}^{4}}{\left(v_{L}^{2}+v_{T}^{2}\right)v_{T}^{4}}\frac{v_{T}l_{T}}{2}\right)
\eea

\section{Discussion}
\label{disc}

\subsection{Approximations used in this work}
\label{discone}
{
There are two approximations that are central to our results: The independent scattering approximation (ISA), and the restriction on the effective wave numbers $K_t,L$ given by
\beq
| Im[K_{L,T}^{2}] | \ll | k^{2}-Re[K_{L,T}^{2}] | \, .
\label{eqoneone}
\eeq
The ISA  means that each dislocation present in the two dimensional elastic medium is characterized by a position and Burgers vector that are statistically independent random variables. This leads to expressions (\ref{WTIisaexpressions})  for the mass operator and for the irreducible kernel, that are linear in the density of dislocations $n$, and involve the single-scattering $T$ matrix. This $T$ matrix has been computed to all orders in a Born approximation expansion.
}

{
The inequality (\ref{eqoneone}) can be satisfied at fixed dislocation density by going to low frequencies (smaller than the renormalized oscillation frequency $\om_{0R}$ defined in Eq. (\ref{def_renorm_freq})), and at fixed frequency by going to low dislocation densities. In the important case of resonance, $\om \sim \om_{0R}$, the restriction (\ref{eqoneone}) can be satisfied with interparticle distances that are small compared to wavelength. }

{
\subsection{Nature of the wave-dislocation interaction and its consequences for coherent wave behavior}
In this paper we have considered the behavior of plane elastic waves in a two dimensional elastic medium in interaction with many edge dislocations that can oscillate around an equilibrium position, with a restoring force given by the Peierls-Nabarro force. The fact that there is a natural frequency associated with the dislocations implies the possibility of resonant scattering. The equation that gives the response to an external loading are solved, and the solution replaced into the wave equation for particle velocity in the presence of dislocations, to give an overall equation for particle velocity only in the presence of a random, frequency dependent, ``potential'', that is of second order in the space derivatives, provided by the dislocations. First, their coherent behavior has been elucidated: there exists average transverse and longitudinal waves, whose phase velocity and attenuation have been explicitly calculated as a function of dislocation density and wave frequency. This has been done solving for the mass operator in a Dyson equation, with the mass operator calculated using an ISA, according to which it is the product of the dislocation density and the averaged single-dislocation $T$-matrix. The latter has been calculated to all orders in a Born approximation scheme. This part of the work follows previous calculations of Maurel et al.~\cite{Maurel2004a}, although the sum to all orders is new.
}

{
\subsection{Nature of the wave-dislocation interaction and its consequences for diffusive wave behavior}
The remaining bulk of this work has been devoted to the study of the diffusion behavior of said waves. The reasoning that has been followed is in accordance with procedures established in other contexts, but the nature of the ``potential'' term introduces unique characteristics. A Bethe-Salpeter equation is written for the irreducible kernel of the two-point field amplitude correlations. Next a Ward-Takahashi identity, needed as a link between the irreducible kernel and the mass operator, is established. This is done following the methods of Barabanenkov~\cite{Barabanenkov1991,Barabanenkov1995}. That is, establishing first a pre-WTI that provides a link between the irreducible kernel and the averaged Green's functions. Although the BS equation is written out in terms of rank-4 tensors, such as the tensor product of two Green's functions, both the pre-WTI and the WTI itself are written in terms of rank-2 tensors. It turns out this is enough to slove the BS equation. In order to establish, or not, diffusive behavior, it is necessary to study the behavior of the BS equation at low frequencies and wavenumbers. As for the coherent waves, an ISA is introduced, and it is verified that this approximation is consistent with the WTI. Next the BS equation is set up, following~\cite{Barabanenkov1991,Barabanenkov1995} as an eigenvalue problem. Eigenfunctions and eigenvalues are found to leading order in low frequency and wavenumber. This portion of the work has a straightforward logic but the algebra is rather cumbersome. At length, a diffusion behavior is established and an explicit expression for the (frequency dependent) diffusion coefficient, Eq. (\ref{DCFintotleadexpl}). is obtained.
}

{
\subsection{Relation of the obtained diffusion behavior to other work}
\label{sec:b=0}
The obtained diffusion coefficient {appears to vanish} when the wave frequency coincides with the resonant frequency of the dislocations. The vanishing of the diffusion constant of electrons, treated as de Broglie waves, in the presence of randomly placed impurities was linked to their (``Anderson'') localization~\cite{Lee1985} and spurred a large body of research into analogous properties of classical (i.e. non-quantum) waves. The role of resonant interaction between classical electromagnetic radiation and matter has been reviewed by Lagendijk and van Tiggelen~\cite{Lagendijk1996}. The effect of resonances in the transport of light in a random system of dielectric spheres has been studied by Sapienza et al.~\cite{Sapienza2007}, by Cowan et al.~\cite{Cowan2011} in the transport of ultrasound in  a system of disordered plastic spheres immersed in water, and by Tallon et al.~\cite{Tallon2020} in emulsions. {In our case, however, since the formalism we have developed holds only in the loss less ($B=0$) case, we cannot set $\om = \om_0$ in Eq. (\ref{prefin}) because it would lead to a singular potential. But we can ask ourselves about the behavior of the problem as $\om$ approaches $\om_0$ {from} below. And the answer to this question is that the diffusion coefficient becomes smaller and smaller, until the behavior determined by internal losses, as encapsulated by the $B$ coefficient, takes over. The role of internal losses in experiments looking for the localization of acoustic waves has been recently discussed by  Cobus et al.~\cite{Cobus2018}.} It will be of interest to see what is the behavior of elastic waves, in three dimensions, in the presence of dislocation segments of finite length, with a resonant frequency in the THz range~\cite{Bianchi2020}. At very low frequency, our diffusion constant becomes a two-dimensional version of the diffusion constant found by Ryzhik et al.~\cite{Ryzhik1996} using energy transfer theory for elastic waves interacting with generic obstacles.
}

\section{Concluding remarks}
{
We have studied the properties of elastic waves in interaction with many, randomly placed, edge dislocations both in the coherent and incoherent regimes. The basic scattering mechanism has the novel feature, compared with the more usual studies of wave propagation in random media, that dislocations are dynamical objects. This leads to properties of the ``potential'' term in the wave equations that are wave-vector dependent, also at variance with previous examples. The present study is part of an ongoing project to explore the consequences of such an interaction for coherent and incoherent wave behavior, both as a matter of principle and in light of possible applications. We have noted in the Introduction that in the coherent, low frequency regime, these studies have led to novel tools for the non-destructive evaluation of the plastic behavior of metals and alloys.
}

{
As noted in the Introduction, there is an increasing amount of experimental evidence pointing to a role played by dislocations in thermal transport, to an extent that has not been fully understood to date. From this point of view, it is important to have as complete an understanding of the elastic wave-dislocation interaction, and of its consequences, as possible. The approach we have taken in this work is to consider dislocations embedded in a continuum elastic solid, and use the tools of continuum solid mechanics. It is of interest to note that, from this point of view, dislocations are just topological line defects (``Volterra dislocations'' \cite{Volterra1907}) that do not need an underlying crystal structure for their existence. This has the consequence that the formalism can be appplied to any solid, not necessaarily crystals, as long as they can be described through continuum mechanics. This approach has led to a renewed understanding of elastic anomalies in glasses~\cite{Lund2015,Bianchi2020}. Recently, the various mechanisms at play in the thermal transport of amorphous materials has been studied in some detail~\cite{Beltukov2018}, indicating that both propagating (coherent) and diffusive (incoherent) modes are present, in various degrees depending on the phonon frequency. In any case, a scattering process appears to be essential, and it does not appear to be completely clear what exactly is doing the scattering. It will be of interest to see if the mechanisms found in the present paper carry over to three dimensions, and the consequences they have, if any, for the thermal transport in glasses.
}

\section{Acknowledgements}
{
This work has been supported by Fondecyt Grant 1191179.
}

\appendix
\section{Averages of rotation in 2D}
\label{Apaver}
In contrast to 3D case, where Euler angles need to be introduced for description of angular averaging \cite{Maurel2005b}, in the 2D case only polar angle $\theta$ determines the corresponding averages.

We define the average of an arbitrary function $f(\theta)$ over the polar angle as
\beq
<f>=\frac{1}{2\pi}\int\limits_{0}^{2\pi}f(\theta)d\theta
\label{averdef}
\eeq
According to combinatorics rules \cite{Ee2017}, the tensor integral, comprised of the product of radial unit $n$-dimensional vectors $\hat{\vec r}$ ($\hat{\vec r}^{2}=1$)
\beq
<\hat{r}^{i_{1}}\cdots\hat{r}^{i_{k}}>_{\hat{\vec r}}=\frac{1}{\Omega^{(n)}}\int d\Omega^{(n)}_{\hat{\vec r}}\hat{r}^{i_{1}}\cdots\hat{r}^{i_{k}}
\label{prodint}
\eeq
with $\Omega^{n}$ as a $n$-dimensional solid angle, is equal to zero when there is an odd number of unit vectors in the product, and reduces to the totally symmetric isotropic tensor $\tilde{{\cal L}}^{i_{1}\cdots i_{2k}}_{(2k)}$ by the rule
\beq
<\hat{r}^{i_{1}}\cdots\hat{r}^{i_{2k}}>_{\hat{\vec r}}=\tilde{{\cal L}}^{i_{1}\cdots i_{2k}}_{(2k)}
\label{prodinttens}
\eeq
with $\tilde{{\cal L}}^{i_{1}\cdots i_{2k}}_{(2k)}$ obeying the recurrence relation
\beq
\tilde{{\cal L}}^{i_{1}\cdots i_{2k}}_{(2k)}=\frac{1}{n+2k-2}
\left(\delta_{i_{1}i_{2}}\tilde{{\cal L}}^{i_{3}\cdots i_{2k}}_{(2k-2)}+\delta_{i_{1}i_{3}}\tilde{{\cal L}}^{i_{2}i_{4}\cdots i_{2k}}_{(2k-2)}+\cdots+\delta_{i_{1}i_{2k}}\tilde{{\cal L}}^{i_{2}\cdots i_{2k-1}}_{(2k-2)}\right)
\label{recurrence}
\eeq
and satisfying to the condition $\tilde{{\cal L}}_{0}=1$.

Recalling that $\Omega^{(2)}=2\pi$, $d\Omega^{(2)}_{\hat{\vec r}}=d\theta$, one can easily see that definitions from Eqn.(\ref{averdef}) and Eqn.(\ref{prodint}) coincides with each other if $f=\hat{r}^{i_{1}}\cdots\hat{r}^{i_{k}}$.

Based on the above, we construct a list of averages of our particular interest
\bea
\mbox{$k=1$:} \hspace{4em} <\hat{r}^{i}\hat{r}^{j}>&=&\frac{\delta_{ij}}{n}\nonumber\\
\mbox{$k=2$:} \hspace{2.4em} <\hat{r}^{i}\hat{r}^{j}\hat{r}^{k}\hat{r}^{l}>&=&\frac{1}{n+2}\left(\delta_{ij}\tilde{{\cal L}}^{kl}_{(2)}+\delta_{ik}\tilde{{\cal L}}^{jl}_{(2)}+\delta_{il}\tilde{{\cal L}}^{jk}_{(2)}\right)\nonumber\\
&=&\frac{1}{n(n+2)}\left(\delta_{ij}\delta_{kl}+\delta_{ik}\delta_{jl}+\delta_{il}\delta_{jk}\right) \, ,
\label{list}
\eea
where $n=2$ in Eqn. (\ref{list}) for two dimensions, the case of interest here.

In addition, since relation from the Eqn. (\ref{unitvectors}) holds in 2D, then, the following important average can be evaluated using Eqn. (\ref{list})
\bea
<\mM_{ab}\mM_{cd}>&=&<(n_a t_b+n_b t_a)(n_c t_d+n_d t_c)> \nonumber \\
&=&<(\epsilon_{ae} t_e t_b+\epsilon_{bf} t_f t_a)(\epsilon_{cg} t_g t_d+\epsilon_{dh} t_h t_c)> \nonumber\\
&=&(\delta_{ac}<t_b t_d>+\delta_{ad}<t_b t_c>+\delta_{bc}<t_a t_d>+\delta_{bd}<t_a t_c>-4<t_a t_b t_c t_d>)\nonumber\\
& =&\frac{\delta_{bc}\delta_{ad}+\delta_{bd}\delta_{ac}-\delta_{ab}\delta_{cd}}{2}
\label{Mprodaver}
\eea

{We shall need this result in the next \ref{Apb}}

\section{Summation of the perturbation expansion for the mass operator}
\label{Apb}
We introduce the definition of the $t$-matrix of $n$-th scatterer in momentum space through (in what follows we avoid frequency argument as a matter of convenience) \cite{Churochkin2016}
\beq
t^{n}_{ab}(\vec k,\vec k')=\int d\vec x d\vec x' e^{-i \vec k \cdot \vec x}t^{n}_{ab}(x,x')e^{i \vec k' \cdot \vec x'}
\eeq
and its Born expansion
\beq
t^{n}_{ab}(\vec k,\vec k')=t^{n(1)}_{ab}(\vec k,\vec k')+t^{n(2)}_{ab}(\vec k,\vec k')+t^{n(3)}_{ab}(\vec k,\vec k') \ldots
\label{tseries}
\eeq
The first Born approximation is specified as:
\bea
t^{n(1)}_{ab}(\vec k,\vec k')&=&\int d\vec x e^{-i \vec k \cdot \vec x}V^{n}_{ab}(x)e^{i \vec k' \cdot \vec x}\nonumber \\
& =&{ \int d\vec x e^{-i \vec k \cdot \vec x}\mu{\cal A}\mM_{ac}^{n}\frac{\partial}{\partial x_c}\delta (\vec x-\vec X_{0}^{n})\mM_{eb}^{n}\frac{\partial}{\partial x_e}|_{\vec x=\vec X_0^{n}}e^{i \vec k' \cdot \vec x}}
\nonumber \\
 & =&{ -\mu{\mathcal A}\mM_{ac}^{n}k_c k'_e \mM_{eb}^{n}e^{i \left(\vec k'-\vec k \right)\cdot \vec X^{n}_0} . }
 \label{TBorn}
\eea
Here,  we used the expression for perturbation potential of a single edge dislocation from Eqn. (\ref{potentialFour}).

The second order Born approximation is given by {
\bea
\label{SecondB}
t^{n(2)}_{ab}(\vec k,\vec k')&=&\int d\vec x \, d\vec x' e^{-i\vec k \cdot \vec x}V^{n}_{ac}(x)G^0_{cd} (x-x') V^{n}_{db} (x') e^{i\vec k' \cdot \vec x'} \\
& = & \frac{(\mu{\mathcal A})^2}{(2\pi)^2} \int d\vec x \, d\vec x'  d \vec q \, \mM_{ac_{1}}^{n} \mM_{c_{2}c}^{n} \mM_{dd_{1}}^{n} \mM_{d_{2}b}^{n}  \nonumber \\
& & \hspace{1em} \times e^{-i\vec k \cdot \vec x} \frac{\partial}{\partial x_{c_{1}}} \delta (\vec x - \vec X^{n}_0) (-iq_{c_{2}}) e^{-i\vec q \cdot \vec X_{0}^{n}} G_{cd}^0(\vec q ) e^{i\vec q \cdot \vec x'} \frac{\partial}{\partial x'_{d_{1}}} \delta (\vec x' - \vec X_{0}^{n}) (ik'_{d_{2}}) e^{i\vec k' \cdot \vec X_{0}^{n}} \nonumber \\
& = & \frac{(\mu{\mathcal A})^2}{(2\pi)^2} \int d\vec q \, \mM_{ac_{1}^{n}} \mM_{c_{2}c}^{n} \mM_{dd_{1}}^{n} \mM_{d_{2}b}^{n} k'_{d_{2}} k_{c_{1}} q_{c_{2}} q_{d_{1}} G_{cd}^0(\vec q )e^{i \left(\vec k'-\vec k \right)\cdot \vec X^{n}_0} \nonumber \\
& = & (\mu{\mathcal A})^2\mM_{ac_{1}}^{n} \mM_{c_{2}c}^{n} \mM_{dd_{1}}^{n} \mM_{d_{2}b}^{n} k'_{d_{2}} k_{c_{1}} e^{i \left(\vec k'-\vec k \right)\cdot \vec X^{n}_0}\frac{1}{2\pi}\int dq \, q^{3}\nonumber \\
&   & \left(\frac{\delta_{cd}}{\rho c_T^2(q^2-  k_T^2)}\frac{1}{2\pi}\int\limits_{0}^{2\pi}\hat q_{c_{2}} \hat q_{d_{1}}d\theta+\left(\frac{1}{\rho c_L^2(q^2-k_L^2)}-\frac{1}{\rho c_T^2(q^2-  k_T^2)}\right) \frac{1}{2\pi}\int\limits_{0}^{2\pi}\hat q_{c_{2}} \hat q_{d_{1}}\hat q_c \hat q_d d\theta\right)\nonumber \\
& = & (\mu{\mathcal A})^2\mM_{ac_{1}}^{n} \mM_{c_{2}c}^{n} \mM_{dd_{1}}^{n} \mM_{d_{2}b}^{n} k'_{d_{2}} k_{c_{1}} e^{i \left(\vec k'-\vec k \right)\cdot \vec X^{n}_0}\frac{1}{2\pi}\int dq \, q^{3}\nonumber \\
&   & \left(\frac{\delta_{cd}\delta_{c_{2}d_{1}}}{2\rho c_T^2(q^2-  k_T^2)}+\left(\frac{1}{\rho c_L^2(q^2-k_L^2)}-\frac{1}{\rho c_T^2(q^2-  k_T^2)}\right) \frac{1}{8}\left(\delta_{c_{2}d_{1}}\delta_{cd}+\delta_{c_{2}c}\delta_{d_{1}d}+\delta_{c_{2}d}\delta_{d_{1}c}\right)\right)\nonumber\\
& = & (\mu{\mathcal A})\mM_{ac_{1}}^{n}\mM_{d_{2}b}^{n} k'_{d_{2}} k_{c_{1}} e^{i \left(\vec k'-\vec k \right)\cdot \vec X^{n}_0}\frac{1}{4\pi}\int dq \, q^{3}\left(\frac{1}{\rho c_T^2(q^2-  k_T^2)}+\frac{1}{\rho c_L^2(q^2-k_L^2)}\right)\nonumber \\
& = & (\mu{\mathcal A})^2\mM_{ac_{1}}^{n}k_{c_{1}} I \mM_{d_{2}b}^{n} k'_{d_{2}}  e^{i \left(\vec k'-\vec k \right)\cdot \vec X^{n}_0}\, , \nonumber
\eea
}
where
\bea
\label{Iint}
I=\frac{1}{4\pi}\int dq \, q^{3}\left(\frac{1}{\rho c_T^2(q^2-  k_T^2)}+\frac{1}{\rho c_L^2(q^2-k_L^2)}\right).
\eea

We notice that, to get the result of Eqn.(\ref{SecondB}) we utilized explicit expressions for the potential and the free Green function from Eqns. (\ref{potentialFour})
and (\ref{g0Sol}), respectively, along with Eqn.(\ref{list}) from \ref{Apaver} and properties of $\mM$:
 \beq
\mM_{mm}=0 \, , \hspace{2em} \mM_{ab}\mM_{ab}=\mM_{ba}\mM_{ba}=2 \, .
\label{tracem}
\eeq.

{
We notice that the integral (\ref{Iint} diverges due to the use of continuum mechanics: there is no intrinsic length scale associated with the elastic medium and all wavelengths, including infinitely short , are included. To regularize this situation we proceed as in \cite{Churochkin2017} and introduce a short distance cut-off $\Lambda$. With this regularization procedure we obtain
\beq
I= \frac{1}{8\pi \rho} \Lambda^2 \left( \frac{1}{c_L^2} + \frac{1}{c_T^2} \right)+ \frac{i}{8\rho} \om^2 \left( \frac{1}{c_L^4} + \frac{1}{c_T^4} \right)
\label{eyefin}
\eeq
}

The third order Born approximation is given by
\beq
t^{n(3)}_{ab}(\vec k,\vec k')    =   \int d\vec x \, d\vec x' d\vec x'' \; e^{-i\vec k \cdot \vec x} V_{ac}(\vec x)  G^0_{cd}(\vec x-\vec x') V_{de}(\vec x') G^0_{ef}(\vec x'-\vec x'')V_{fb}(\vec x'') e^{i\vec k' \cdot \vec x''} \, .
\label{third term}
\eeq
Proceeding in the same way as for the second order yields:
{
\beq
{t^{n(3)}_{ab}(\vec k,\vec k')   = -(\mu{\mathcal A})^3\mM_{ac}k_{c} I^2 \mM_{db}^{n} k'_{d}  e^{i \left(\vec k'-\vec k \right)\cdot \vec X^{n}_0}\, .}
\label{third term2}
\eeq

Based on Eqns.(\ref{TBorn}, \ref{SecondB}, \ref{third term2}), summation of the corresponding geometrical series for $t^{n}_{ab}(\vec k,\vec k')$ can be easily performed with the result
\bea\label{rest}
t^{n}_{ab}(\vec k,\vec k') &=& t^{n(1)}_{ab}(\vec k,\vec k')+t^{n(2)}_{ab}(\vec k,\vec k')+t^{n(3)}_{ab}(\vec k,\vec k')+....\\ &=& { \frac{-\mu{\mathcal A}}{1+\mu{\mathcal A}I}\mM_{ac}^{n}k_{c}\mM_{db}^{n} k'_{d} e^{i \left(\vec k'-\vec k \right)\cdot \vec X^{n}_0}\,} \nonumber.
\eea
Given that Eqns.(\ref{massISA}, \ref{rest}, \ref{Mprodaver}), we have for the mass operator
{
\bea\label{mo}
\Sigma_{ab}(\vec k)&=& n \int <t^{n}_{ab}(\vec k,\vec k')> d\vec X^{n}_0\\
&=&-n\mu\frac{{\mathcal A}}{1+\mu{\mathcal A}I}k_{c}k_{d}\frac{\left(\delta_{ad}\delta_{cb}+\delta_{ab}\delta_{cd}-\delta_{ac}\delta_{db}\right)}{2}\nonumber\\
&=&-n\mu\frac{1}{2}\frac{{\mathcal A}}{1+\mu{\mathcal A}I}k^{2}\delta_{ab}\nonumber  \hspace{2em} \\
&=& -\frac{n\mu^2 b^2}{2M} \frac{k^2 \delta_{ab}}{\om^2 -\om^2_{0R} +i (\om B/M + \om^2 \kappa^2)}
\label{massopweye}
\eea}
}
{where
\beq
\label{def_renorm_freq}
\om_{0R}^2 \equiv \om_0^2 - \frac{\mu b^2}{8\pi M} \left( \frac{c_T^2}{c_L^2} + 1 \right) \Lambda^2
\eeq
is a renormalized frequency of dislocation oscillation about its equilibrium position, and
\beq
\kappa^2 \equiv \frac{\mu b^2}{8Mc_T^2}  \left( \frac{c_T^4}{c_L^4} + 1 \right) \, .
\eeq
is a dimensionless constant.
In arriving at (\ref{massopweye}) we have used (\ref{eyefin}).
}

\section{Optical theorem}
\label{OT}
Here we show that Eqn. (\ref{WTIeventzero})
\bea
\left(\Sigma^{\ast}_{bc}({\bf k})-\Sigma_{bc}({\bf k})\right) = {\int\limits_{\mathbf{k}_1}}\left(G^{0\ast}(\vec k_1)_{gh}-G^{0}(\vec k_1)_{gh}\right)
K_{gh,bc}({\bf k}_1,{\bf k})
\nonumber
\eea
holds, in the ISA, { when $B=0$} . That is, when the mass and irreducible vertex operators are given in terms of the $t$ matrix by (\ref{WTIisaexpressions}), and the $t$ matrix itself is given by (\ref{rest}). Consider the left-hand-side first:
\bea
\left(\Sigma^{\ast}_{bc}({\bf k})-\Sigma_{bc}({\bf k})\right)  &=& Im\left[\frac{n\mu{\mathcal A}}{1+\mu{\mathcal A}I}\right]k^{2}\delta_{dc}   \\
&=& Im\left[\frac{n\mu{\mathcal A}\left(1+\mu{\mathcal A}I^{\ast}\right)}{\left(1+\mu{\mathcal A}I\right)\left(1+\mu{\mathcal A}I^{\ast}\right)}\right] \\
&=& -\left[\frac{n\mu^2{\mathcal A}^2}{\left(1+\mu{\mathcal A}I\right)\left(1+\mu{\mathcal A}I^{\ast}\right)}\right] Im[I] \\
&=&  -\left[\frac{n\mu^2{\mathcal A}^2}{\left(1+\mu{\mathcal A}I\right)\left(1+\mu{\mathcal A}I^{\ast}\right)}\right] Im \left[\frac{1}{4\pi}\int dq \, q^{3}\left(\frac{1}{\rho c_T^2(q^2-  k_T^2)}+\frac{1}{\rho c_L^2(q^2-k_L^2)}\right) \right] \nonumber \\
&=& -\left[\frac{n\mu^2{\mathcal A}^2}{\left(1+\mu{\mathcal A}I\right)\left(1+\mu{\mathcal A}I^{\ast}\right)}\right] \frac{1}{4}\int dq \, q^{3}\left(\frac{\delta\left(q^{2}-k^{2}_{T}\right)}{\rho c^{2}_{T}}+\frac{\delta\left(q^{2}-k^{2}_{L}\right)}{\rho c^{2}_{L}}\right)
\label{lhsc14}
\eea

And now the right-hand-side:

\bea
\label{proof}
{\int\limits_{\mathbf{k}_1}}\left(G^{0\ast}(\vec k_1)_{gh}-G^{0}(\vec k_1)_{gh}\right)
K_{gh,bc}({\bf k}_1,{\bf k})&  & \\
&& \hspace{-14em} = -\frac{n}{2\pi}\int\limits_{k_{1}}\int\limits_{\hat {\bf k}_{1}}k^{3}_{1}\left(\frac{\delta\left(k^{2}_{1}-k^{2}_{T}\right)}{\rho c^{2}_{T}}\delta_{ab}\hat k_{1g} \hat k_{1f}+\left(\frac{\delta\left(k^{2}_{1}-k^{2}_{L}\right)}{\rho c^{2}_{L}}-\frac{\delta\left(k^{2}_{1}-k^{2}_{T}\right)}{\rho c^{2}_{T}}\right)\hat k_{1a} \hat k_{1b}\hat k_{1g} \hat k_{1f}\right)\nonumber\\
&& \hspace{-7em}\times \frac{\mu{\mathcal A}}{1+\mu{\mathcal A}I}\left(\frac{\mu{\mathcal A}}{1+\mu{\mathcal A}I}\right)^{\ast} <\mM_{ag}^{n}\mM_{fb}^{n}\mM_{hc}^{n}\mM_{de}^{n} > k_{h}k_{e} \nonumber\\
&&\hspace{-14em} = -n\int\limits_{k_{1}}k^{3}_{1}\left(\frac{\delta\left(k^{2}_{1}-k^{2}_{T}\right)}{\rho c^{2}_{T}}\delta_{ab}\frac{\delta_{gf}}{2}+\left(\frac{\delta\left(k^{2}_{1}-k^{2}_{L}\right)}{\rho c^{2}_{L}}-\frac{\delta\left(k^{2}_{1}-k^{2}_{T}\right)}{\rho c^{2}_{T}}\right)\frac{\delta_{ab}\delta_{gf}+\delta_{ag}\delta_{bf}+\delta_{af}\delta_{bg}}{8}\right)\nonumber\\
&& \hspace{-7em}\times \frac{\mu{\mathcal A}}{1+\mu{\mathcal A}I}\left(\frac{\mu{\mathcal A}}{1+\mu{\mathcal A}I}\right)^{\ast} <\mM_{ag}^{n}\mM_{fb}^{n}\mM_{hc}^{n}\mM_{de}^{n}>k_{h}k_{e} \nonumber  \\ \vspace{1em}
&& \hspace{-14em} = -n\int\limits_{k_{1}}\frac{k^{3}_{1}}{2}\left(\frac{\delta\left(k^{2}_{1}-k^{2}_{T}\right)}{\rho c^{2}_{T}}+\frac{\delta\left(k^{2}_{1}-k^{2}_{L}\right)}{\rho c^{2}_{L}}\right)
\frac{\mu{\mathcal A}}{1+\mu{\mathcal A}I}\left(\frac{\mu{\mathcal A}}{1+\mu{\mathcal A}I}\right)^{\ast}<\mM_{hc}^{n}\mM_{de}^{n}>k_{h}k_{e}
\eea
which, using (\ref{Mprodaver}) is easily seen to be equal to (\ref{lhsc14}).
\section{Solution  for $f^{1\mathbf{q}}({\bf k})$}
\label{solu}
To find $f^{1\mathbf{q}}({\bf k})$ we need (\ref{system1q}). Using (\ref{zeroorder}) its second term can be rewritten as
\begin{eqnarray}\label{ApSolTran}
\int\limits_{\bf{k}^{\prime\prime}}H^{1\mathbf{q}}({\bf k},{\bf k}^{\prime\prime}){\cal{B}}\Delta G({\bf k}^{\prime\prime}) & = &
\int\limits_{\bf{k}^{\prime\prime}}{\cal{B}}\left(P(\mathbf{q};{\bf k}^{\prime\prime})\delta_{\mathbf{k}^{\prime\prime},\mathbf{k}}\Delta G({\bf k}) - H({\bf k},{\bf k}^{\prime\prime})\Delta G^{1\mathbf{q}}({\bf k}^{\prime\prime})\right) \nonumber \\
 & = & {\cal{B}}\int\limits_{\bf{k}^{\prime\prime}}H({\bf k},{\bf k}^{\prime\prime})
 [ \int\limits_{{\bf k}_{2}}\Phi({\bf k}^{\prime\prime},{\bf k}_{2})P(\mathbf{q};{\bf k}_{2})- \Delta G^{1\mathbf{q}}({\bf k}^{\prime\prime}) ]\nonumber \, .
\end{eqnarray}
Where the first equality is a consequence of the symmetry property from Eqn. (\ref{symmetry})  and application the WTI to $H^{1\mathbf{q}}({\bf k},{\bf k}^{\prime\prime})$. The second equality is a result of substituting $\delta_{\mathbf{k}^{\prime\prime},\mathbf{k}}\Delta G({\bf k})$ by its value given by (\ref{BSfinal}).
Hence,
\beq\label{ApSol1q}
f^{1\mathbf{q}}({\bf k}^{\prime\prime})=
-{\cal{B}}\left(\int\limits_{{\bf k}_{2}}\Phi({\bf k}^{\prime\prime},{\bf k}_{2})P(\mathbf{q};{\bf k}_{2})-\Delta G^{1\mathbf{q}}({\bf k}^{\prime\prime})\right)
\eeq
with
\bea\label{ApSoldeltaG}
\Delta G^{1\mathbf{q}}_{cd,ee}({\bf k})&=&{\bf q}\cdot\frac{\partial\Delta G_{cd,ee}({\bf k};{\bf q}^{\prime},0)}{\partial {\bf q}^{\prime}}|_{{\bf q}^{\prime}=0}\\ &=& -\frac{1}{2\imath\rho}{\bf q}\cdot\frac{\partial \left(Re[G_{cd}(\mathbf{k})]\right)}{\partial {\bf k}} \\
&=& -\frac{q_{h}Re[G_{L}-G_{T}]}{2\imath\rho}\frac{\partial P_{\textbf{\^{k}}}}{\partial k_{h}}
-\frac{q_{h}}{2\imath\rho}\left(\frac{\partial \left(Re[G_{T}]\right)}{\partial k_{h}}\left(\textbf{I}-P_{\textbf{\^{k}}}\right)+
\frac{\partial \left(Re[G_{L}]\right)}{\partial k_{h}}P_{\textbf{\^{k}}}\right)
\eea
and
\begin{eqnarray}\label{ApSolDerivative}
\frac{\partial P_{\textbf{\^{k}}}}{\partial k_{h}}&=&\frac{\partial \left(\frac{k_{c}k_{d}}{k^{2}}\right)}{\partial k_{h}}=\left(\frac{k_{d}\delta_{ch}+k_{c}\delta_{dh}}{k^{2}}\right)-\frac{2k_{c}k_{d}k_{h}}{k^{4}}\\
Re[G_{T,L}]&=&\frac{F_{T,L}(\omega,k)}{\rho\omega^{2}Im[K_{T,L}^{2}]}
\left(Re[K_{T,L}^{2}]\left(k^2-Re[K_{T,L}^{2}]\right)-Im[K_{T,L}^{2}]^{2}\right)\\
\frac{\partial \left(Re[G_{T,L}]\right)}{\partial k_{g}}&=&\frac{2 k_{g}F_{T,L}(\omega,k)}{\rho\omega^{2}Im[K_{T,L}^{2}]}
\left(2k^{2}F_{T,L}(\omega,k)Im[K_{T,L}^{2}]-Re[K_{T,L}^{2}]\right)\\
F_{T,L}(\omega,k)&=&\left(\frac{Im[K_{T,L}^{2}]}{\left(k^2-Re[K_{T,L}^{2}]\right)^{2}+Im[K_{T,L}^{2}]^{2}}\right)
\label{ApSolDerivative1}
\end{eqnarray}
\section{Calculation of $D_{\Delta G^{1\mathbf{q}}}$}
\label{esti}
From Eqn. (\ref{Dconstant}) we have
\begin{eqnarray}\label{Gq}
D_{\Delta G^{1\mathbf{q}}} & = &
\frac{-{\cal{B}}^{2}}{q^{2}\omega\left(1+a\right)}
\int\limits_{\bf{k}}P_{aa,cd}({\bf k};{\bf q})
\Delta G_{cd,ee}^{1\mathbf{q}}({\bf k})\\
& = &\frac{{\cal{B}}^{2}}{4\rho^{2}q^{2}\omega\left(1+a\right)}
\int\limits_{\bf{k}}q_{h}\frac{\partial L_{dc}({\bf k})}{\partial k_{h}}
\frac{\partial \left(Re[G^{-}_{cd}(\mathbf{k})]\right)}{\partial k_{g}}q_{g}\nonumber\\
& = &\frac{-{\cal{B}}^{2}q_{h}q_{g}\left(c_{L}^{2}-c_{T}^{2}\right)}{2\rho q^{2}\omega\left(1+a\right)}\int\limits_{\bf{k}}
\left(Re[G_{L}-G_{T}]\left(\delta_{hg}-\frac{k_{h}k_{g}}{k^{2}}\right)\right)\nonumber\\
&  & \hspace{2em}+\frac{-B^{2}q_{h}q_{g}}{2\rho q^{2}\omega\left(1+a\right)}\int\limits_{\bf{k}}
\left(c_{T}^{2}\frac{\partial \left(Re[G_{T}]\right)}{\partial k_{g}}+c_{L}^{2}\frac{\partial \left(Re[G_{L}]\right)}{\partial k_{g}}\right)k_{h}\nonumber
\end{eqnarray}
where Eqn. (\ref{Gq}) has been obtained using Eqns. (\ref{parameters}) and (\ref{ApSoldeltaG}, \ref{ApSolDerivative}).

In Eqn. (\ref{Gq}) we have to deal with the following two types of integrals
\begin{eqnarray}\label{ApEstInt}
\mathbb{I}^{hg}_{T,L}&=&\int\limits_{\bf{k}}
Re[G_{T,L}]\left(\delta_{hg}-\frac{k_{h}k_{g}}{k^{2}}\right)=\frac{\delta_{hg}}{4\pi}\int\limits_{-\infty}^{\infty}k\Theta(k) Re[G_{T,L}]dk\\
\mathbb{J}^{hg}_{T,L}&=&
\int\limits_{\bf{k}}
\left(\frac{\partial \left(Re[G_{T,L}]\right)}{\partial k_{g}}\right)k_{h}\nonumber
\end{eqnarray}
Then, using Eqn. (\ref{ApSolDerivative}) we can write
\begin{eqnarray}\label{ApEstInt2}
\mathbb{I}^{hg}_{T,L}& = &\int\limits_{-\infty}^{\infty} \frac{\delta_{hg}k\Theta(k)F_{T,L}(\omega,k)\left(Re[K_{T,L}^{2}]\left(k^2-Re[K_{T,L}^{2}]\right)-Im[K_{T,L}^{2}]^{2}\right)}{4\pi\rho\omega^{2}Im[K_{T,L}^{2}]}dk\\
\mathbb{J}^{hg}_{T,L}& = & \int\limits_{\bf{k}}
\frac{2 k_{g}k_{h}\left(2k^{2}F_{T,L}^{2}(\omega,k)Im[K_{T,L}^{2}]-Re[K_{T,L}^{2}]F_{T,L}(\omega,k)\right)}{\rho\omega^{2}Im[K_{T,L}^{2}]}\nonumber\\
& = & \int\limits_{-\infty}^{\infty}\delta_{gh}\Theta(k)k^{3}dk
\left(\frac{ 2k^{2}F_{T,L}^{2}(\omega,k)Im[K_{T,L}^{2}]-Re[K_{T,L}^{2}]F_{T,L}(\omega,k)}{2\pi\rho\omega^{2}Im[K_{T,L}^{2}]}
\right).\nonumber
\end{eqnarray}
 The expression for $\mathbb{J}^{hg}_{T,L}$ in Eqn. (\ref{ApEstInt2}) contains an ill-defined term in the integrand, proportional to $F_{T,L}^{2}(\omega,k)$, that can be regularized as follows \cite{Mahan2000}: Introduce a new variable $x_{T,L} \equiv \left(k^2-Re[K_{T,L}^{2}]\right)$, and consider the following auxiliary integrals:
\begin{eqnarray}\label{AuxInt}
\int\limits_{-\infty}^{\infty}\left(\frac{Im[K_{T,L}^{2}]}{x_{T,L}^{2}+Im[K_{T,L}^{2}]^{2}}\right)dx_{T,L}&=&\pi,\\
\int\limits_{-\infty}^{\infty}\left(\frac{Im[K_{T,L}^{2}]}{x_{T,L}^{2}+Im[K_{T,L}^{2}]^{2}}\right)^{2}dx_{T,L}&=&\frac{\pi}{2Im[K_{T,L}^{2}]}.\nonumber
\end{eqnarray}
They show that it is possible to make the replacements
\begin{eqnarray}\label{Rep}
\left(\frac{Im[K_{T,L}^{2}]}{x_{T,L}^{2}+Im[K_{T,L}^{2}]^{2}}\right) & \longrightarrow & \pi\delta(x_{T,L}),\\
\left(\frac{Im[K_{T,L}^{2}]}{x_{T,L}^{2}+Im[K_{T,L}^{2}]^{2}}\right)^{2} & \longrightarrow & \frac{\pi\delta(x_{T,L})}{2Im[K_{T,L}^{2}]}.\nonumber
\end{eqnarray}
which yield the same result after integration over $x_{T,L}$. Moreover, both replacements give the proper asymptotic behaviour in the low-density limit (our case) when $Im[K_{T,L}^{2}]\rightarrow 0$  {or, more precisely, when $| Im[K_{T,L}^{2}] | \ll | k^{2}-Re[K_{T,L}^{2}] |$}. Therefore, on the basis of Eqns. (\ref{ApSolDerivative}, \ref{Rep}) we have %
\begin{eqnarray}\label{FReg}
F_{T,L}(\omega,k) & = & \pi\delta\left(k^2-Re[K_{T,L}^{2}]\right),\\
F_{T,L}(\omega,k)^{2} & = & \frac{\pi\delta\left(k^2-Re[K_{T,L}^{2}]\right)}{2Im[K_{T,L}^{2}]}\nonumber
\end{eqnarray}
Technically, Eqn. (\ref{FReg}) indicates that in all final expressions the limit ${Im[K_{T,L}^{2}]\rightarrow 0}$. {The implications of this restriction are discussed in Section \ref{discone}}.

From Eqns. (\ref{Gq}), (\ref{ApEstInt}), (\ref{ApEstInt2}), and (\ref{FReg}) it follows that
\begin{eqnarray}\label{Dqeval}
\mathbb{I}^{hg}_{T,L}& = & \frac{-\delta_{hg}Im[K_{T,L}^{2}]}{8\rho\omega^{2}},\quad
\mathbb{J}^{hg}=0\Rightarrow\nonumber\\
D_{\Delta G^{1\mathbf{q}}}& = &
\frac{{\cal{B}}^{2}\left(c_{L}^{2}-c_{T}^{2}\right)\left(Im[K_{L}^{2}]-Im[K_{T}^{2}]\right))}{16\rho^{2} \omega^{3}\left(1+a\right)}\, .
\end{eqnarray}
\section{Improved diffusion constant for the case of screw dislocations}
\label{WTIscrewimpr}
The application of the technique for WTI derivation, which is presented in this paper, to the case of screw dislocations requires respective modification of the result~\cite{Churochkin2017} for WTI, and, consequently, diffusion constant. Indeed, given that Eqns. (6, 7, 11) from~\cite{Churochkin2017}, for screw dislocations, the system of equations similar to Eqns. (\ref{greeneqI}, \ref{greeneqII}) can be written as
\bea
\rho\omega_{1}^{2} G(\vec x_{1},\vec x^{\prime}_{1},\omega_{1})+\mu\nabla^{2}_{1} G(\vec x_{1},\vec x^{\prime}_{1},\omega_{1})& =&-\mu V(\vec x_{1},\omega_{1})G(\vec x_{1},\vec x^{\prime}_{1},\omega_{1})-\delta(\vec x_{1}-\vec x^{\prime}_{1})
\label{screwgreeneqI} \\
\rho\omega_{2}^{2} G(\vec x_{2},\vec x^{\prime}_{2},\omega_{2})+\mu\nabla^{2}_{2}G(\vec x_{2},\vec x^{\prime}_{2},\omega_{2})& =& -\mu V(\vec x_{2},\omega_{2})G(\vec x_{2},\vec x^{\prime}_{2},\omega_{2})-\delta(\vec x_{2}-\vec x^{\prime}_{2}) \, .
\label{screwgreeneqII}
\eea
with
\beq\label{screwpotanti}
\left. V({\bf x},\omega)={\cal A}\sum\limits_{n=1}^{N} \;
 \frac{\partial}{\partial x_a}  \delta ( \vec x-\vec X^{n}_0 )\;
{\frac{\partial}{\partial x_a}} \right|_{\vec x=\vec X^{n}_0}
\eeq
and
 \beq\label{screwprefanti}
{\cal A}=\frac{\mu b^2}{M} \frac{1}{\omega^2 {+} i \omega (B/M) -\omega^2_0 }=\frac{\mu b^2}{M}g(\omega)
\eeq
where $g(\omega) \equiv \frac{1}{\omega^2 {+} i \omega (B/M) -\omega^2_0 }$. Subscripts $1,2$ at $\nabla^{2}_{1,2}$ indicate action of operator $\nabla^{2}$  on components of vectors $\vec x_{1}$ and $\vec x_{2}$, respectively. Again, we consider all dislocations have a Burgers vector of the same magnitude, but possibly different sign. Finally, we explicitly introduced $g(\omega)$ here, which is formally the same as for edge dislocations (see notation right after Eqn. (\ref{greeneqIift})), but it is completely different from  $g(\omega)$ for screw dislocations that was introduced in~\cite{Churochkin2017}.

In fact, the way of the potential subtraction is the reason for such difference. Namely, in~\cite{Churochkin2017}, we initially integrated out the potential (see Eqns. (45, 46) in~\cite{Churochkin2017}) instead of following the substraction procedure, that is outlined in Section \ref{sec:pre-WTI} for the case of edge dislocations and getting somewhat similar to Eqn. (\ref{preWTI}). The former approach is independent of a particular form of frequency dependence of the potential, i. e., $g(\omega)$ is reduced to $\omega^{2}$, which, in fact, is a frequency dependence of the free medium Green function, whereas the latter approach is still capturing a particular frequency dependence of the potential, so that $g(\omega)$ is identified through it. Therefore, the second approach is considered to be more precise, say differentiated, whereas the first one is integral and, as a consequence of that, it is losing a crucially important information that specifies frequency dependence of a particular potential. And this Appendix is aimed to get an improved WTI and, consequently, diffusion constant within the second approach adapted for the case of screw dislocations.

The adaptation is essentially passing through steps similar to Eqns. (\ref{greeneqIift}-\ref{preWTI}), but with system of Eqns. (\ref{screwgreeneqI}, \ref{screwgreeneqII}) that includes a scalar potential $\mu V(\vec x,\omega)$ and Green function $G$. To be more specific, with definitions from Eqns. (\ref{ift}, \ref{dft}) used in Eqns. (\ref{screwgreeneqI}, \ref{screwgreeneqII}) one can get
\bea\label{screwgreeneqsysift}
-(G^{0})^{-1}(\vec k_{1},\omega_{1})\delta_{\vec k_{1},\vec k^{\prime\prime}_{1}}g^{*}(\omega_{2})+g^{*}(\omega_{2})(G)^{-1}(\vec k_{1},\vec k^{\prime\prime}_{1};\omega_{1}&=&    \\
&&\hspace{-3em} g(\omega_{1})g^{*}(\omega_{2})\sum\limits_{n=1}^{N}\frac{\mu^2 b^2}{M} k_{h_1}k^{\prime\prime}_{h_1}e^{i \left(\vec k^{\prime\prime}_{1}-\vec k_{1} \right)\cdot \vec X^{n}_0} \nonumber \\
-(G^{0*})^{-1}(\vec k_{2},\omega_{2})\delta_{\vec k_{2},\vec k^{\prime\prime}_{2}}g(\omega_{1})+g(\omega_{1})(G^{*})^{-1}(\vec k_{2},\vec k^{\prime\prime}_{2};\omega_{2})&=&  \\
&& \hspace{-3em} g(\omega_{1})g^{*}(\omega_{2})\sum\limits_{n=1}^{N}\frac{\mu^2 b^2}{M} k_{h_2}k^{\prime\prime}_{h_2}e^{-i \left(\vec k^{\prime\prime}_{2}-\vec k_{2} \right)\cdot \vec X^{n}_0} \nonumber
\eea
with
\bea
(G^{0})^{-1}(\vec k,\omega)=-\left(\rho\omega^{2}-\mu k^{2}\right).
 \label{screwg0ft}
\eea
It should be noted that Eqns. (\ref{screwgreeneqsysift}) is formally a scalar version of Eqns. (\ref{greeneqsysift}). Then, the next step is the substraction at $B=0$ ($g(\omega)=g^{*}(\omega)$) of both equations from each other keeping the following limit:
$\vec k_{2}\rightarrow\vec k^{\prime\prime}_{1}$, $\vec k^{\prime\prime}_{2}\rightarrow\vec k_{1}$. This results in the expression
\bea\label{screwpreWTIft}
\lim\limits_{{\substack{\vec k_{2}\rightarrow\vec k^{\prime\prime}_{1}\\ \vec k^{\prime\prime}_{2}\rightarrow \vec k_{1}}}}
\left(-(G^{0})^{-1}(\vec k_{1},\omega_{1})\delta_{\vec k_{1},\vec k^{\prime\prime}_{1}}g^{*}(\omega_{2})+g^{*}(\omega_{2})(G)^{-1}(\vec k_{1},\vec k^{\prime\prime}_{1};\omega_{1})+\right.&&\nonumber\\ \left.(G^{*0})^{-1}(\vec k_{2},\omega_{2})\delta_{\vec k_{2},\vec k^{\prime\prime}_{2}}g(\omega_{1})-g(\omega_{1})(G^{*})^{-1}(\vec k_{2},\vec k^{\prime\prime}_{2};\omega_{2})\right)&\equiv&0\nonumber
\eea
Again, by acting on the identity with
$\lim\limits_{{\substack{\vec k_{2}\rightarrow\vec k^{\prime\prime}_{1}\\ \vec k^{\prime\prime}_{2}\rightarrow \vec k_{1}}}} G(\vec k^{\prime\prime}_{1},\vec k^{\prime\prime\prime}_{1} ;\omega_{1})G^{*}(\vec k^{\prime\prime}_{2},\vec k^{\prime\prime\prime}_{2};\omega_{2})$
from the right, averaging out the random variables, and introducing notations from Eqn. (\ref{notat}), we get
\bea\label{screwpreWTIftprod}
-<G^{-}(\vec k^{\prime\prime\prime-},\vec k^{-} ;\omega_{-})(G^{0+})^{-1}(\vec k^{+},\omega_{+})G^{+}(\vec k^{+},\vec k^{\prime\prime\prime+} ;\omega_{+})>g^{\ast}(\omega_{-}) \hspace{2em}&& \nonumber \\  + g^{\ast}(\omega_{-})<\delta_{\vec k,\vec k^{\prime\prime\prime}}G^{-}(\vec k^{-},\vec k^{\prime\prime\prime-};\omega_{-})>&&\nonumber\\ +<G^{-}(\vec k^{\prime\prime\prime-},\vec k^{\prime\prime-};\omega_{-})(G^{0-})^{-1}(\vec k^{\prime\prime-},\omega_{-})G^{+}(\vec k^{\prime\prime+},\vec k^{\prime\prime\prime+} ;\omega_{+})g(\omega_{+})> \hspace{2em}&& \nonumber \\
-g(\omega_{+})<\delta_{\vec k^{\prime\prime},\vec k^{\prime\prime\prime}}G^{+}(\vec k^{\prime\prime+},\vec k^{\prime\prime\prime+} ;\omega_{+})>&\equiv& 0
\eea
Eventually, this expression provides us with a final form for pre-WTI
\bea\label{screwpreWTI}
\left((G^{0-})^{-1}({\bf k};{\bf q},\Omega)g(\omega_{+})-(G^{0+})^{-1}({\bf k};{\bf q},\Omega)g^{\ast}(\omega_{-})\right)\Phi({\bf k},{\bf k}^{\prime\prime\prime};{\bf q},\Omega)+&&\\ g^{\ast}(\omega_{-})<G>^{-}(\vec k^{\prime\prime\prime};{\bf q},\Omega)-g(\omega_{+})<G>^{+}(\vec k^{\prime\prime\prime};{\bf q},\Omega)&\equiv&0,\nonumber
\eea
with notations
\begin{eqnarray}\label{screwintensityft}
(G^{0\pm})^{-1}(\vec k^{\pm},\omega_{\pm})&=&(G^{0\pm})^{-1}({\bf k};{\bf q},\Omega) \\
<G^{\pm}(\vec k^{\prime\prime\prime\pm},\vec k^{\prime\prime\prime\pm} ;\omega_{\pm})>&=&<G>^{\pm}(\vec k^{\prime\prime\prime};{\bf q},\Omega)\nonumber\\
\Phi({\bf k},{\bf k}^{\prime};{\bf q},\Omega)&\equiv &<G^{+}(\mathbf{k}^{+},\mathbf{k}^{\prime\prime\prime +},\omega^{+})G^{-}(\mathbf{k}^{\prime\prime\prime-},\mathbf{k}^{-},\omega^{-})>\nonumber
\end{eqnarray}
The form of pre-WTI for screw dislocations presented by Eqn. (\ref{screwpreWTI}}) is formally considered to be a scalar version of pre-WTI for edge dislocations Eqn. (\ref{preWTI}}) and would have reduced to the case of pre-WTI for screw dislocations published in~\cite{Churochkin2017} if $g(\omega)$ had been equal to $\omega^{2}$. This connection suggests the following logic to get an expression for diffusion constant $D$: Scalar versions of Eqns. (\ref{WLWTI}, \ref{Dconstant}) for WTI and diffusion constant for the case of edge dislocations, are identical to corresponding relations published in~\cite{Churochkin2017} (see, Eqns. (54, 79) therein), but with a new $g(\omega)$ presented in this paper, and, consequently, with a new $a$ defined by a scalar version of Eqn. (\ref{aparameter}). Consequently, the diffusion constant can be explicitly written as follows
\begin{eqnarray}\label{screwDconstant}
D &\equiv & -\frac{i\Omega\lambda^{2\mathbf{q}}}{q^2\lambda^{1\Omega}}=-\frac{{\cal{B}}}{q^{2}\omega\left(1+a\right)}
\int\limits_{\bf{k}}P({\bf k};{\bf q})f^{1\mathbf{q}}({\bf k}) \\
 &=&
\frac{{\cal{B}}^{2}}{q^{2}\omega\left(1+a\right)}
\int\limits_{\bf{k}}P({\bf k};{\bf q})
\left(\int\limits_{{\bf k}_{2}}\Phi({\bf k},{\bf k}_{2})P(\mathbf{q};{\bf k}_{2})-\Delta G^{1\mathbf{q}}({\bf k})\right) \label{screwDconstant2} \\
&\equiv& D^{\mathcal{R}} + D_{\Delta G^{1\mathbf{q}}}
\label{screwDconstant3}
\end{eqnarray}
with
\begin{eqnarray}
\label{screwaparameter}
a&=&\frac{1}{\int\limits_{\bf{k}}f({\bf k})}\int\limits_{\bf{k}^{\prime\prime}} \frac{\left(A(\mathbf{k}^{\prime\prime};\mathbf{0},\Omega)\left(g(\omega_{+})-g^{\ast}(\omega_{-})\right)\right)^{1\Omega}}{2\omega\Omega}f({\bf k}^{\prime\prime}) \\
g(\omega) &\equiv&\left. \frac{1}{\omega^2 {+} i \omega (B/M) -\omega^2_0 }\right|_{B=0}.
\end{eqnarray}
The definition of remaining entities in Eqns. (\ref{screwDconstant}-\ref{screwaparameter}) is the same as in~\cite{Churochkin2017}. It means that the replacement of $g(\omega)$ does not affect the form of results for $D^{\mathcal{R}}$, $D_{\Delta G^{1\mathbf{q}}}$ up to replacement of $a$, providing us with a final result for diffusion constant of screw dislocations
\beq\label{screwDCFin}
D =
\frac{{\cal{B}}^{2}}{q^{2}\omega\left(1+a\right)}
\int\limits_{\bf{k}}P({\bf k};{\bf q})
G({\bf k})G^{*}({\bf k})P({\bf k};{\bf q})
\eeq
{In order to analyze the modified result for screw dislocations we have to find the explicit expression for $D$. It is established on the basis of the result for the diffusion constant (we denote it as $D_{Isc}$) from ~\cite{Churochkin2017}, which reads as
\beq\label{screwDCI}
D_{Isc}=
\frac{{\cal{B}}^{2}}{q^{2}\omega\left(1+a_{Isc}\right)}
\int\limits_{\bf{k}}P({\bf k};{\bf q})
G({\bf k})G^{*}({\bf k})P({\bf k};{\bf q})
\eeq
where $a_{Isc}$ is the corresponding version of $a$ parameter. In other words,
\beq\label{screwDrelation}
D=\frac{\left(1+a_{Isc}\right)}{\left(1+a\right)}D_{Isc}
\eeq
According to Eqn.(95) from~\cite{Churochkin2017}, we have
\beq\label{screwDCIexpl}
D_{Isc}=\frac{c_{T}^{4}}{2\omega^{3}\left(1+a_{Isc}\right)}\frac{K^{2}(K^{2})^{*}}{Im[K^{2}]}\approx
\frac{c_{T}^{4}}{v^{4}\left(1+a_{Isc}\right)}\frac{vl}{2}\approx \frac{c_{T}^{2}}{v^{2}}\frac{vl}{2}
\eeq
Substituting Eqn.(\ref{screwDCIexpl}) into Eqn.(\ref{screwDrelation}) we come to the result
\beq\label{screwDrelationexpl}
D=\frac{c_{T}^{4}}{2\omega^{3}\left(1+a\right)}\frac{K^{2}(K^{2})^{*}}{Im[K^{2}]}\approx
\frac{c_{T}^{4}}{v^{4}\left(1+a\right)}\frac{vl}{2}
\eeq
with $a$ from Eqn. (\ref{screwaparameter}). Based on Eqn. (\ref{screwaparameter}) and Eqn. (76) from~\cite{Churochkin2017} it can be proven the following relation
\beq\label{arelat}
a=\frac{\omega^{2}}{\omega^{2}_{0}-\omega^{2}}a_{Isc}
\eeq
This provides us with the approximation (see Eqn.(77) in~\cite{Churochkin2017})
\beq\label{arelatapprox}
a\approx\frac{\omega^{2}}{\omega^{2}_{0}-\omega^{2}}\left(\frac{c_{T}^{2}Re[K^{2}]}{\omega^{2}}-1\right)\approx\frac{\left(\frac{c_{T}^{2}}{v^{2}}-1\right)}{\frac{\omega^{2}_{0}}{\omega^{2}}-1}
\eeq
Eventually, we get for $D$
\beq\label{screwDrelationexpl}
D\approx
\frac{c_{T}^{4}}{v^{4}}\frac{1}{\left(1+\frac{\left(\frac{c_{T}^{2}}{v^{2}}-1\right)}{\frac{\omega^{2}_{0}}{\omega^{2}}-1}\right)}\frac{vl}{2}
\eeq
}

There are two specific instances of interest for this result: The first is that $D=0$ when $\om = \om_0$, raising the possibility of elastic wave localization by scattering from dislocations. {As discussed in Section \ref{sec:b=0}, however, a strict equality cannot hold because in that case the scattering problem becomes singular. But the diffusion constant does decrease as the frequency approaches the resonance frequency from below.} The second is  the low frequency, or long wavelength, limit. Now, there are two length scales to compare the wavelength $\lambda=c_T/\om$ against: the mean distance between dislocations, $1/\sqrt{n}$, and the Peierls-Nabarro frequency divided by the speed of shear waves $c_T/\om_0$. When $\lambda$ is small compared to both, use of Eqs. (21) and (24) of \cite{Churochkin2017} leads to (when $B=0$, which is the case for which we have derived a WTI)
\beq
\lim_{\om \to 0} D = \frac 12 c_T\ell
\eeq
with
\beq
\ell = \frac{\alpha}{n} \frac{\om}{c_T} \frac{\om_{0R}^4}{\om^4}
\eeq
with $\alpha = \left(\ln (\delta/\epsilon)\right)^{2}/2\pi^{2}$ a dimensionless constant of order one. This is of the form $\ell \sim 1/n\sigma$ where $\sigma \propto \om^3$ is a scattering cross section corresponding to Rayleigh scattering in two dimensions.
}



%

\end{document}